\documentclass[12pt, a4paper]{article}
\pdfoutput=1
\usepackage{graphicx}
\usepackage{amssymb}
\usepackage{amsmath}
\usepackage{bm}
\usepackage{color}
\usepackage{theorem}
\usepackage{colortbl}
\usepackage{subcaption}

\usepackage[sort&compress,numbers, merge]{natbib}

\definecolor{Orange}{cmyk}{0,0.61,0.87,0}
\definecolor{JungleGreen}{cmyk}{0.99,0,0.52,0}
\definecolor{OliveGreen}{cmyk}{0.64,0,0.95,0.40}
\definecolor{Brown}{cmyk}{0,0.70,1,0.40}
\definecolor{RoyalBlue}{cmyk}{0.71,0.53,0,0.12}
\definecolor{Gray}{cmyk}{0,0,0,0.40}
\definecolor{LightPink}{cmyk}{0.0,0.25,0,0}
\definecolor{LLightPink}{cmyk}{0.0,0.10,0,0}
\definecolor{LightBlue}{cmyk}{0.25,0,0,0}
\definecolor{LightGray}{cmyk}{0,0,0,0.2}

\allowdisplaybreaks[1]

\usepackage[colorlinks=true, linkcolor=OliveGreen, citecolor=RoyalBlue,
urlcolor=Brown]{hyperref}

\newcommand{\beq}{\begin{equation}}
\newcommand{\eeq}{\end{equation}}
\newcommand{\bea}{\begin{eqnarray}}
\newcommand{\eea}{\end{eqnarray}}

\begin{document}

\begin{titlepage}

\begin{flushright}
FTPI--MINN--25/07,
UMN--TH--4501/25, \\
KCL-PH-TH/2025-36, CERN-TH-2025-160
\end{flushright}

\vskip 1cm
\begin{center}

{\LARGE
{\bf
Deformations of Starobinsky Inflation in \\
\vspace{3mm} No-Scale SU(5) and SO(10) GUTs
}
}

\vskip 1cm

{\bf John Ellis}$^{a,b}$,
{\bf Marcos A. G. Garc{\'i}a}$^{c}$, 
{\bf Natsumi Nagata}$^{d}$,\\ 
\vspace{1.5mm}
{\bf Dimitri V. Nanopoulos}$^{b,e,f}$
and {\bf Keith A. Olive}$^{g}$

\vskip 0.6cm

{\it $^a$Theoretical Particle Physics and Cosmology Group, Department of
  Physics, King's~College~London, London WC2R 2LS, United Kingdom}\\
  {\it $^b$Theoretical Physics Department, CERN, CH-1211 Geneva 23,
  Switzerland}\\
  {\it $^c$Departamento de F\'isica Te\'orica, Instituto de F\'isica, Universidad Nacional Aut\'onoma de M\'exico, Ciudad de M\'exico C.P. 04510, Mexico}\\
  {\it $^d$Department of Physics, University of Tokyo, Bunkyo-ku, Tokyo
 113--0033, Japan}\\ 
 {\it $^e$Academy of Athens, Division of Natural Sciences, Athens 10679, Greece}\\
 {\it $^f$George P. and Cynthia W. Mitchell Institute for Fundamental Physics and Astronomy, Texas A\&M University, College Station, TX 77843, USA}\\
  {\it $^g$William I.~Fine Theoretical Physics Institute, School of Physics and Astronomy, University of Minnesota, Minneapolis, MN 55455, USA}\\

\date{\today}

\vskip 0.5cm

\begin{abstract}
The original Starobinsky $R + R^2$ model of inflation is consistent with Planck and other measurements of the CMB, but recent results from the ACT and SPT Collaborations hint that the tilt of scalar perturbations may be in tension with the prediction of the Starobinsky model. No-scale models of inflation can reproduce the predictions of the Starobinsky model, but also provide a framework for incorporating deformations that could accommodate more easily the ACT and SPT data. We discuss this possibility in the contexts of SU(5) GUTs, taking into account the constraints on these models imposed by the longevity of the proton, the cold dark matter density and the measured value of the Higgs boson. We find that SU(5) with a CMSSM-like pattern of soft supersymmetry breaking has difficulty in accommodating all the constraints, whereas SU(5) with pure gravity-mediated supersymmetry breaking can accommodate them easily. We also consider two SO(10) symmetry-breaking patterns that can accommodate the ACT and SPT data. In both the SU(5) and SO(10) models, the deformations avoid issues associated with large initial field values in the Starobinsky model: in particular, the total number of e-folds is largely independent of the initial conditions.

\end{abstract}

\end{center}
\end{titlepage}

\section{Introduction}

Historically, the first model of cosmological inflation was the $R + R^2$ deformation of the minimal Einstein-Hilbert action for general relativity proposed by Starobinsky~\cite{Staro}. Subsequently many field-theoretical models of inflation have been proposed \cite{reviews}. In parallel, there have been many experimental measurements, notably by the Planck satellite~\cite{Planck}, of the tilt of scalar perturbations, $n_s$, and upper limits \cite{rlimit,Tristram:2021tvh} on the tensor-to-scalar perturbation ratio, $r$, which have been constraining progressively models of inflation. Until recently, the measurements by Planck and terrestrial experiments have been consistent with the predictions of the original Starobinsky model while excluding many alternatives. This trend has focused theoretical attention on field-theoretical models that reproduce, at least approximately, the Starobinsky predictions. These include no-scale models of inflation~\cite{Ellis:2020lnc} and their generalizations such as $\alpha$-attractor models \cite{Kallosh:2013yoa,enov3,enov4}.

Recently, two terrestrial experiments have reported new results pertaining to the cosmological observables. The Atacama Cosmology Telescope (ACT)  Collaboration~\cite{ACT:2025fju,ACT:2025tim} has reported values of $n_s$ that are in some tension with the Starobinsky prediction. In some contrast, the South Pole Telescope (SPT)-3G Collaboration~\cite{SPT-3G:2025bzu} has reported results that are in relatively good agreement with Planck.  The results from ACT have stimulated many theoretical studies of field-theoretical models that allow deformations of the effective inflationary potential in the Starobinsky model~\cite{Kallosh:2025rni,Aoki:2025wld,Dioguardi:2025vci,Antoniadis:2025pfa, German:2025mzg, He:2025bli,Drees:2025ngb,Zharov:2025evb,Addazi:2025qra,Maity:2025czp,Mondal:2025kur,Haque:2025uri,Liu:2025qca,Gialamas:2025ofz,Yogesh:2025wak,Yi:2025dms,Saini:2025jlc,Haque:2025uga, Wang:2025dbj, Hai:2025wvs,Heidarian:2025drk,Wolf:2025ecy,Leontaris:2025hly,Pallis:2025gii,German:2025ide}. In this paper we revisit one such possibility, namely that heavy GUT Higgs fields may modify the effective inflationary potential at large inflaton field values, increasing the value of $n_s$~\cite{Ellis:2020lnc}.

The ACT and SPT results provide hints for addressing one of the challenges for inflationary model building, namely the embedding of the inflation model in a UV extension of the Standard Model such as a GUT or string theory. We have long argued that $N=1$ supersymmetry renders more natural the formulation of a realistic field-theoretical inflationary model \cite{Cries}, just as it addresses the naturalness problem of the Standard Model~\cite{Maiani:1979cx}. However, embedding inflation in $N=1$ supergravity is non-trivial, as the supergravity Lagrangian typically mixes different sectors of the theory. This may either affect GUT symmetry breaking or modify the inflationary dynamics. The ACT and SPT data may be suggesting that the latter is a feature, rather than a bug.

Previously, we have considered combinations of a no-scale supergravity \cite{no-scale} formulation of the Starobinsky model \cite{eno6,eno7} with SO(10) \cite{egnno1} and flipped SU(5) $\times$ U(1) \cite{egnno2} GUT models. Related studies based on a free-fermion formulation of string theory were considered in \cite{anr1,ano}. Here our primary focus is to study simple SU(5) GUTs and update previous work on an SO(10) GUT. We pointed out in~\cite{Ellis:2020lnc}, see Eq.~(216), that the Starobinsky prediction for $n_s$ would in general be subject to a correction related to the vacuum expectation value (VEV) of the ${\bf 24}$ Higgs in the SU(5) GUT, and used the Planck data to set an upper limit on this deformation of the Starobinsky model. The ACT data, in particular, may now be suggesting that this correction is non-zero. 

However, this possibility should not be considered in isolation, but in the broader context of SU(5) GUT phenomenology, including the proton lifetime, the cold dark matter density $\Omega_{\rm CDM} h^2$ and the Standard Model Higgs mass, $m_h$. In this paper we show that ACT, SPT and these other constraints are difficult to reconcile within a variant of the SU(5) GUT with (quasi-)universal soft supersymmetry breaking at the GUT scale, but that reconciliation is possible if pure gravity mediation (PGM) \cite{pgm} is assumed for soft supersymmetry breaking. In the case of the SO(10) GUTs, the two models we study can both
accommodate any of the Planck, ACT or SPT data, subject to the appropriate model constraints being satisfied.

An interesting common feature of the deformations of the Starobinsky potential in both SU(5) and SO(10) GUTs is that they raise it at large field values, eliminating the characteristic Starobinsky plateau. This has the effect of shielding the models from difficulties associated with large initial field values, e.g., the total number of e-folds is largely independent of the initial conditions---in sharp contrast to the Starobinsky model, where the number of e-folds increases
exponentially with the initial field value \cite{Antoniadis:2025pfa}. The total number of e-folds can be interpreted as
a measure of the probability that our patch of the Universe originated from a given
initial state, with larger numbers of e-folds corresponding to higher probabilities. In the models studied here,
large initial field values yield similar numbers of e-folds and hence may be interpreted
as equally probable, alleviating one of the main objections to Starobinsky-like inflationary models.

The layout of this paper is as follows. In Section~\ref{sec:noscalStaro} we review the basic features of Starobinsky-like inflation in no-scale supergravity. In Section~\ref{sec:data} we review the status of the CMB data, with particular emphasis on the interpretation of the recent results from ACT and SPT. In Section~\ref{sec:SU5} we discuss the status of SU(5) GUT models in light of the ACT and SPT data, and in Section~\ref{sec:SO10} we give an update on the SO(10) GUT model. Finally, Section~\ref{sec:conx} summarizes our conclusions.

\section{No-Scale Starobinsky Basics}
\label{sec:noscalStaro}

We set the scene by reviewing briefly the construction of Starobinsky and Starobinsky-like inflation models in the context of no-scale supergravity~\cite{no-scale}. Obtaining the Starobinsky potential along a real direction in field space requires at least two chiral superfields~\cite{eno7} that parametrize
a non-compact SU(2,1)/SU(2)$\times$U(1) coset space.
The corresponding K\"ahler potential
can be written as
\begin{equation}
K \; \ni \; - 3 \, \ln \left(T + T^* - \frac{|\phi|^2}{3} + \dots \right) + \dots \, , 
\label{K21}
\end{equation}
where $T$ and $\phi$ are complex scalar fields. Reheating requires a coupling of the inflaton to Standard Model fields, which are represented by the $\dots$ in Eq.~(\ref{K21}). These are ``untwisted" if they appear in the argument of the logarithm, ``twisted" if outside. In this parameterization and in the context of string compactification, the field $T$ can be interpreted as a volume modulus.
Restricting our attention to the two-field case for the moment,
the minimal version of the K\"ahler potential (\ref{K21}) yields the following kinetic terms
for the scalar fields $T$ and $\phi$:
\begin{equation}
{\cal L}_{\rm KE} =   
\frac{3}{(T + T^* - |\phi|^2/3)^2} ~
\left( \partial_\mu \phi^*, \partial_\mu T^* \right)
\begin{pmatrix} 
(T + T^*)/3 & - \phi \\ - \phi^* & 1 
\end{pmatrix}
\begin{pmatrix} 
\partial^\mu \phi \\ \partial^\mu T 
\end{pmatrix}  \, .
\label{no-scaleL}
\end{equation}
For a general superpotential $W(T,\phi)$, the
effective potential becomes
\begin{equation}
V \; = \; \frac{{\hat V}}{(T + T^* - |\phi|^2/3)^2}  ~,
\label{VVhatT}
\eeq
with
\beq
{\hat V} \; \equiv \; \left| \frac{\partial W}{\partial \phi} \right|^2  +\frac{1}{3} (T+T^*) |W_T|^2 +
\frac{1}{3} \left(W_T (\phi^* W_\phi^* - 3 W^*) + {\rm h.c.}  \right) \, ,
\label{effV}
\end{equation}
where $W_\phi \equiv \partial W/\partial \phi$ and $W_T \equiv \partial
W/\partial T$. 

There are many possible choices for the superpotential \cite{eno7,enov1} that lead to the Starobinsky potential \cite{Staro} 
\begin{equation} 
V =  \frac{3}{4} M^2 (1- e^{-\sqrt{2/3}x})^2 \, ,
\label{r2pot}
\end{equation} 
or similar potentials, 
where $x$ is the canonically-normalized inflaton field. One simple choice for $W$ is \cite{eno6}
\begin{equation}
W \; = \; \left(\frac{M}{2} \phi^2 - \frac{\lambda_\phi}{3\sqrt{3}} \phi^3 \right) \,  .
\label{WZW}
\end{equation}
If
 $T$ is fixed at its vacuum expectation value (VEV) taken for convenience to be $2
\,\langle {\rm Re} T \rangle = 1$ and $\langle {\rm Im} T \rangle = 0$,
the canonically-normalized field along the real $\phi$ direction is
\beq
\phi = \sqrt{3} \tanh \left(x/\sqrt{6}\right) \, ,
\label{canphi}
\eeq
and the scalar potential along the real $x$ direction is identical to the Starobinsky inflationary potential \cite{eno6}
if $\lambda_\phi = M/M_P$,  with $M_P^{-2} = 8 \pi G_N \equiv 1$. The VEV of $T$ can be arbitrary so long as it is stabilized. There are several ways to achieve stabilization, e.g., by adding terms of higher order in $(T+T^*)^n$, with $n\ge4$ \cite{EKN3,eno7,EGNO4}. Alternatively, these fields can be stabilized by including a dilaton contribution to the K\"ahler potential of the form $-\log(S+S^*)$ \cite{ano}. We do not discuss the issue of stabilization here, but refer the reader to~\cite{Ellis:2020lnc} for more detail. 

In an alternative construction, the volume modulus $T$ can be identified with the inflaton if the field
$\phi$ is fixed with $\phi = 0$, and the superpotential is given by
\cite{Cecotti} 
\begin{equation}
W \; = \sqrt{3} M \phi (T - 1/2) \, .
\label{W3}
\end{equation}
The Starobinsky potential (\ref{r2pot}) is found when $T$ is converted
to a canonical field via
\beq
T = \frac12 e^{\sqrt{2/3}x}
\, .
\eeq
As discussed in~\cite{eno7,enov1}, there is a large class of superpotentials
that all lead to the same inflationary potential, but we do not go into details here. 

For a suitably large initial value of $x$, there is sufficient cosmological expansion for inflation to resolve the standard set of cosmological problems \cite{reviews}. 
The slow-roll inflationary parameters are given by
\begin{equation} 
\epsilon \; \equiv \; \frac{1}{2} M_{P}^2 \left( \frac{V'}{V} \right)^2 ; \; \;  \eta \; \equiv \; M_{P}^2 \left( \frac{V''}{V} \right)   \, ,
\label{epsilon}
\end{equation}
where a prime denotes a derivative with respect to the canonically-normalized inflaton field, $x$. The inflationary observables are then calculated in terms of these slow-roll parameters. The scalar tilt, $n_s$, can be approximated by 
\beq
n_s \;  \simeq \; 1 - 6 \epsilon_* + 2 \eta_* \, ,
\label{ns}
\eeq
where the $_*$ subscript denotes a suitable pivot scale. The experimental value at the pivot scale $k_* = 0.05$~Mpc$^{-1}$ extracted
from Planck data~\cite{Planck} in combination with gravitational lensing is
\beq
n_s \; = \; 0.9649 \pm 0.0042 \; (68\%~{\rm CL}) \, 
\label{nsexp}
\eeq
Alternative estimates including recent measurements by the ACT~\cite{ACT:2025fju,ACT:2025tim} and SPT~\cite{SPT-3G:2025bzu}
experiments are discussed below.  
In addition to $n_s$, inflation models are constrained by the scalar-to-tensor ratio:
\beq
r \; 
\simeq  \; 16 \epsilon_* \label{r}
\eeq
for which the current 95\% CL upper bound on $r$ is $0.036$~\cite{rlimit,Tristram:2021tvh}.
In addition, we have the amplitude of the scalar spectrum given by 
\beq\label{eq:As}
A_{S*} \;\simeq\; \frac{V_*}{24\pi^2\epsilon_* M_P^4}\,.
\eeq

The number of e-folds starting from the pivot scale can also be computed from the slow-roll parameters:
\begin{equation}
N_* \;\equiv\; \ln\left(\frac{a_{\rm{end}}}{a_*}\right) \; = \; \int_{t_*}^{t_{\rm{end}}} H dt \; \simeq \;  - \int^{\phi_{\rm{end}}}_{\phi_*} \frac{1}{\sqrt{2 \epsilon_*}} \frac{d \phi}{M_P} \, ,
\label{e-folds}
\end{equation}
where $a_*$ is the value of the cosmological scale factor at the pivot scale, $a_{\rm end}$ is its value at the end of inflation (when exponential expansion ends and $\ddot{a}=0$) and $\phi_*$ and $\phi_{\rm end}$ are the corresponding values of $\phi$ 
at those two epochs. 
In practice, Eq.~(\ref{e-folds}) is used to determine $\phi_*$,
which is subsequently used to determine the slow-roll parameters. The number of e-folds, assuming no additional entropy production after reheating, is given by \cite{LiddleLeach,Martin:2010kz,EGNO5,egnov}:
\begin{equation}
\begin{aligned}
\label{eq:nstarreh}
N_{*} \; &= \; \ln \left[\frac{1}{\sqrt{3}}\left(\frac{\pi^{2}}{30}\right)^{1 / 4}\left(\frac{43}{11}\right)^{1 / 3} \frac{T_{0}}{H_{0}}\right]-\ln \left(\frac{k_{*}}{a_{0} H_{0}}\right) 
-\frac{1}{12} \ln g_{\mathrm{RH}} \\
&+\frac{1}{4} \ln \left(\frac{V_{*}^{2}}{M_{P}^{4} \rho_{\mathrm{end}}}\right) +\frac{1-3 w_{\mathrm{int}}}{12\left(1+w_{\mathrm{int}}\right)} \ln \left(\frac{\rho_{\mathrm{R}}}{\rho_{\text {end }}}\right) 
\, ,
\end{aligned}
\end{equation}
where the present best-fit value of the Hubble parameter found by the Planck Collaboration is $H_0 = 67.36 \, \rm{km \, s^{-1} \, Mpc^{-1}}$ \cite{Planck} and the present photon temperature is $T_0 = 2.7255 \, \rm{K}$~\cite{Fixsen:2009ug}. Here, $\rho_{\rm{end}} \equiv \rho(\phi_{\rm end})$ and $\rho_{\rm{R}}$ are the energy density at the end of inflation and at the beginning of the radiation domination era when $w = p/\rho = 1/3$, respectively, $a_0 = 1$ is the present day scale factor, $g_{\rm{RH}} =  915/4$ is the effective number of relativistic degrees of freedom in the minimal supersymmetric standard model (MSSM) at the time of reheating, and $w_{\rm int}$ is the equation-of-state parameter averaged over the e-folds during reheating. Using
 $k_* \; = \; 0.05 \, \rm{Mpc}^{-1}$, the first line in (\ref{eq:nstarreh}) gives
$N_*  \simeq  61.04$. As a result,
$N_*$ depends on the reheating temperature through $\rho_{\rm R}$ and hence also the inflationary observables, most notably $n_s$.  

Finally, the constant $M$ is determined from the amplitude of density fluctuation in the microwave background radiation $A_s = 2.1 \times 10^{-9}$ \cite{Planck}. In the model (\ref{WZW}) with $\lambda_\phi = 1$, this is given by
\beq
A_s   =   \frac{3 M^2}{8\pi^2} \sinh^4 (\phi_*/\sqrt{6}) \, , \label{As2}
\eeq
and fixes $M  \approx 1.2 \times 10^{-5}$ in Planck
units.

In order to achieve reheating the inflaton must be coupled  to matter. 
In no-scale models, supergravity couplings of the inflaton are strongly
suppressed \cite{ekoty}, and require either a non-trivial coupling
through the gauge kinetic function \cite{ekoty,klor,EGNO4,EGNO5}, or a direct
coupling to the matter sector through the superpotential. It was
proposed in Ref.~\cite{ENO8} that the inflaton could be associated with
the scalar component of the right-handed (SU(2)-singlet) neutrino
superfield $\nu_R$, and a specific no-scale supersymmetric GUT
\cite{superGUT,Ellis:2016tjc} model based on SU(5) was proposed, in
which the $\nu_R$ appeared as a singlet. In this model, reheating takes
place when the inflaton decays into the left-handed sneutrino and Higgs
(or neutrino and Higgsino), and may occur simultaneously with
leptogenesis \cite{lepto}.  

\section{Review of Recent Data}
\label{sec:data}

As mentioned above, a fit to the combination of Planck and lensing data yielded
\begin{equation}
\label{Plens}
n_s \; = \; 0.9649 \pm 0.0042 \; {\rm at~the} \; 68\% \; {\rm CL} \, ,
\end{equation}
which became $n_s = 0.9652 \pm 0.0042$ when combined with BICEP/Keck data~\cite{rlimit}.
As mentioned in the Introduction, two new high-multipole CMB measurements have recently been reported, by the ACT~\cite{ACT:2025fju,ACT:2025tim} and SPT~\cite{SPT-3G:2025bzu} Collaborations.
The ACT measurement by itself is quite compatible with (\ref{Plens}):
$n_s \; = \; 0.9666 \pm 0.0077 \; {\rm at~the} \; 68\% \; {\rm CL}$.
However, because of opposite correlations between $n_s$ and $\Omega_b h^2$ in the Planck and ACT measurements, a fit to the combination of Planck, lensing and ACT data (P-ACT-L) yields a higher value, namely $n_s \; = \; 0.9713 \pm 0.0037 \; {\rm at~the} \; 68\% \; {\rm CL}$,
in some tension with the Planck and lensing combination (\ref{Plens}). This tension is further increased in a fit (P-ACT-LB) where DESI measurements of baryon acoustic oscillations (BAO)~\cite{DESI:2024uvr} are combined with the Planck, lensing and ACT data:
\begin{equation}
\label{PLACTD}
n_s \; = \; 0.9743 \pm 0.0034 \; {\rm at~the} \; 68\% \; {\rm CL} \, .
\end{equation}
More recently, SPT-3G measurements~\cite{SPT-3G:2025bzu} have yielded $n_s = 0.951 \pm 0.011 \; {\rm at~the} \; 68\% \; {\rm CL}$, i.e., a central value somewhat below the Planck/lensing value (\ref{Plens}) but with a larger uncertainty. Combining the Planck and ACT data with theirs, the SPT Collaboration has reported
\begin{equation}
\label{CMB-SPA}
n_s \; = \; 0.9684 \pm 0.0030 \; {\rm at~the} \; 68\% \; {\rm CL} \, .
\end{equation}
In this paper, we consider the implications of the results (\ref{Plens}, \ref{PLACTD}) and (\ref{CMB-SPA}) for possible deformations of Starobinsky inflation in no-scale GUT models.

\section{SU(5) GUTs and No-Scale Inflation}
\label{sec:SU5}

In minimal supersymmetric SU(5), the  three generations of SM quarks and leptons are embedded in ${\bf 10}$ and $\overline{\bf 5}$ representations, $\Psi_i$ and $\Phi_i$, respectively, where $i$ is the generation index, while the MSSM Higgs fields, $H_u$ and $H_d$, reside in ${\bf 5}$ and $\overline{\bf 5}$ representations, $H$ and $\overline{H}$, respectively. The SU(5) GUT symmetry is spontaneously broken by a VEV of a ${\bf 24}$ representation, $\Sigma$, down to the SM gauge group. In addition to these fields, we introduce an SU(5) singlet field, $\phi$, as the inflaton. We also assume that this model respects $R$-parity, so as to suppress dangerous baryon/lepton-number violating renormalizable operators; $H$, $\overline{H}$, $\Sigma$, and $\phi$ are $R$-parity even and the rest of the fields are $R$-parity odd.

The renormalizable superpotential for this model is given by 
\begin{align}
  W_5 &=  \mu_\Sigma {\rm Tr}\Sigma^2 + \frac{1}{6} \lambda^\prime {\rm
  Tr} \Sigma^3 + \mu_H \overline{H} H + \lambda \overline{H} \Sigma H
 \nonumber \\
 &+ \left(h_{\bf 10}\right) 
  \Psi \Psi H +
  \left(h_{\overline{\bf 5}}\right) \Psi \Phi
  \overline{H}
  \nonumber \\ 
  & + \frac{M}{2} \phi^2 - \frac{\lambda_\phi}{3\sqrt{3}} \phi^3 
  + \lambda_{H\phi} \phi \overline{H} H + \lambda_{\phi \Sigma} \phi {\rm Tr}\Sigma^2
  ~,
 \label{W5}
\end{align}
where we have suppressed the tensor structure
and omitted generation indices, for simplicity.
If we choose $\lambda_\phi = M/M_P$, we obtain the Starobinsky potential~\cite{eno6} for $\lambda_{\phi \Sigma} \ll 1$ as we assume in the following. Once SU(5) is broken, the GUT gauge bosons obtain masses $M_X = 5 g_5 V_\Sigma$, where $g_5$ is the SU(5) gauge coupling and $\langle \Sigma \rangle = V_{\Sigma}\, {\rm diag} (2,2,2,-3,-3)$ is the VEV of $\Sigma$, with $V_{\Sigma} \equiv 4\mu_{\Sigma}/\lambda^\prime$.  To realize doublet-triplet mass splitting for the $H$ and $\overline{H}$ multiplets, we must take $\mu_H - 3\lambda V_{\Sigma} \ll V_\Sigma$. In this case, the colored-triplet Higgs bosons have masses $M_{H_C} = 5 \lambda V_\Sigma$.  

The soft supersymmetry-breaking terms in the minimal supersymmetric SU(5) GUT are
\begin{align}
 {\cal L}_{\rm soft} = &- \left(m_{\bf 10}^2\right)_{ij}
 \widetilde{\psi}_i^* \widetilde{\psi}_j
- \left(m_{\overline{\bf 5}}^2\right)_{ij} \widetilde{\phi}^*_i
 \widetilde{\phi}_j
- m_H^2 |H|^2 -m_{\overline{H}}^2 |\overline{H}|^2 - m_\Sigma^2 {\rm Tr}
\left(\Sigma^\dagger \Sigma\right)
\nonumber \\
&-\biggl[
\frac{1}{2}M_5 \widetilde{\lambda}^{A} \widetilde{\lambda}^A
+ A_{\bf 10} \left(h_{\bf 10}\right)_{ij}
 \epsilon_{\alpha\beta\gamma\delta\zeta} \widetilde{\psi}_i^{\alpha\beta}
 \widetilde{\psi}^{\gamma\delta}_j H^\zeta
+ A_{\overline{\bf 5}}\left(h_{\overline{\bf 5}}\right)_{ij}
 \widetilde{\psi}_i^{\alpha\beta} \widetilde{\phi}_{j \alpha}  \overline{H}_\beta
\nonumber \\
&+ B_\Sigma \mu_\Sigma {\rm Tr} \Sigma^2 +\frac{1}{6} A_{\lambda^\prime
 } \lambda^\prime  {\rm Tr} \Sigma^3 +B_H \mu_H \overline{H} H+
 A_\lambda \lambda \overline{H} \Sigma H +{\rm h.c.}
 \biggr]~,
\end{align}
where $\widetilde{\psi}_i$ and $\widetilde{\phi}_i$ are the scalar
components of $\Psi_i$ and $\Phi_i$, respectively,
the $\widetilde{\lambda}^A$ are the SU(5) gauginos, and we
use the same symbols for the scalar components of the Higgs fields as for the
corresponding superfields.

The K\"{a}hler potential includes the inflaton field as an untwisted field. The rest of the fields can be included in the K\"{a}hler potential as either untwisted or twisted fields. 
For concreteness, we assume all of the fields are untwisted in the following discussion.

All of the above fields except $\phi$ and $\Sigma$ have vanishing field values
in the instantaneous potential minimum during inflation. We assume that the adjoint Higgs field is displaced by a small amount from its vacuum value during the inflationary period, i.e., 
\begin{equation}
   \Sigma  = (V_{\Sigma} +\sigma) \, {\rm diag} (2,2,2,-3,-3)  ~,
\end{equation}
with $|\sigma| \ll V_{\Sigma} $. We show below that this condition can be satisfied for sufficiently small $\lambda_{\phi\Sigma}$. In this case, the scalar potential during inflation is given by 
\begin{align}
  V &= \frac{\hat{V}}{(T+T^* -|\phi|^2/3 - 10 |V_\Sigma+\sigma |^2 )^2} 
   ~,
\end{align}
with
\begin{align}
  \hat{V} &= \left|M \phi - \frac{\lambda_\phi}{\sqrt{3}} \phi^2 + 30 \lambda_{\phi\Sigma} (V_\Sigma +\sigma )^2 \right|^2  \nonumber \\
  &+ \frac{15}{2} \left| V_\Sigma +\sigma\right|^2 
  \left|4(\mu_\Sigma + \lambda_{\phi\Sigma}\phi) - \lambda^\prime (V_\Sigma +\sigma )  \right|^2   ~.
\end{align}
The instantaneous value of $\Sigma$ during inflation is determined by the second term in the above equation for $\lambda_{\phi\Sigma} \ll 1$: 
\begin{equation}
  \sigma \simeq \frac{4 \lambda_{\phi\Sigma}\phi}{\lambda^\prime}  ~.
\end{equation}
To ensure $|\sigma| \ll V_{\Sigma}$, we thus assume $\lambda_{\phi\Sigma} \ll \mu_\Sigma / M_P $. This condition is generically weaker than the limit obtained from the bounds on the inflation parameters, as we see below. 

\begin{figure}[ht!]
    \centering
    \includegraphics[width=.53\textwidth]{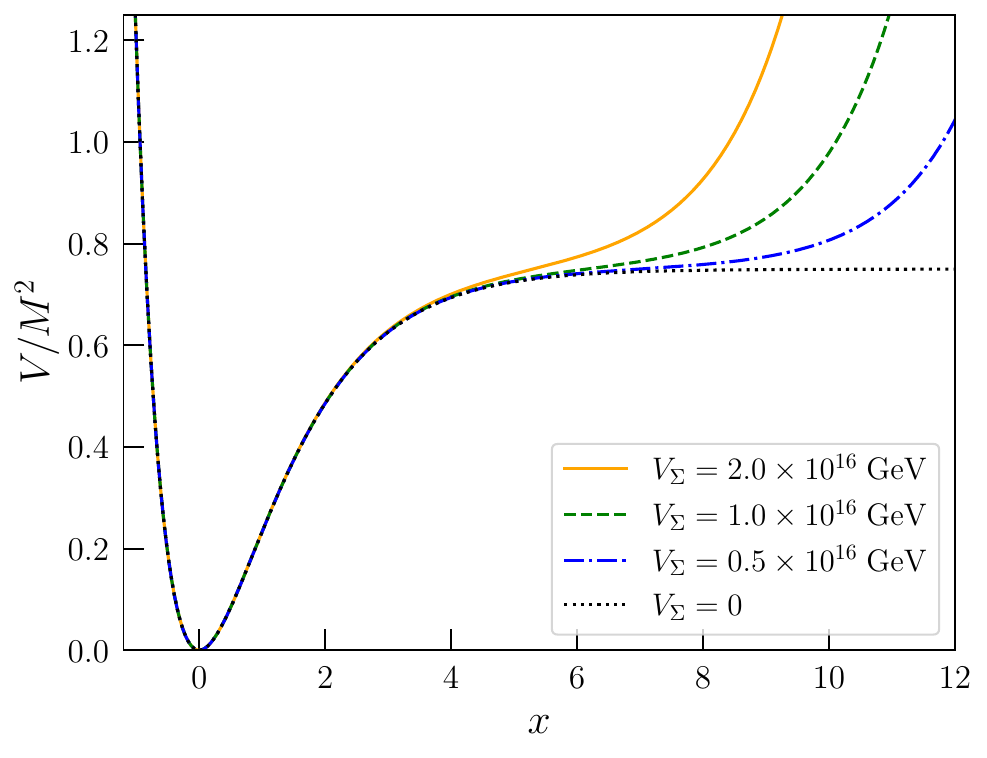}
    \caption{ The Starobinsky potential (black dotted) and its modification in an SU(5) GUT (\ref{eq:delvsu5}) for indicative values of $V_\Sigma$ with $\lambda_{\phi\Sigma} = 0$.}
    \label{staro}
\end{figure}

With $\langle T \rangle = 1/2$ and the canonically-normalized field $x$ given by Eq.~(\ref{canphi}),
the inflaton potential can be approximated by 
\begin{align}
  V \simeq \frac{3}{4} M^2 \left(1-e^{- \sqrt{\frac{2}{3}}x}\right)^2 
  + \Delta V ~,
  \label{delv}
\end{align}
where 
\begin{align}
  \Delta V &= 15 M^2 V_\Sigma^2 e^{- \sqrt{\frac{2}{3}} x} 
  \sinh^2 \biggl(\sqrt{\frac{2}{3}}x\biggr) \nonumber \\
 & + 60 \sqrt{3} \lambda_{\phi\Sigma}\, M V_\Sigma^2 
  e^{-\frac{1}{\sqrt{6}}x} \cosh^2 \biggl(\frac{x}{\sqrt{6}}\biggr)
  \sinh \biggl(\frac{x}{\sqrt{6}}\biggr) \nonumber \\ 
  &\simeq \frac{3}{4} M^2 V_{\Sigma}^2\biggl(
  5 + \frac{30 \lambda_{\phi\Sigma} }{\sqrt{3}M}   
  \biggr) e^{\sqrt{\frac{2}{3}}x} ~. 
  \label{eq:delvsu5}
\end{align}
The CMB observables then take the values 
\begin{align}
  n_s &\simeq 1 - \frac{2}{N_*} + \frac{320}{27} \biggl(\frac{V_\Sigma}{M_P}\biggr)^2 \biggl(
    1 + \frac{2 \sqrt{3}\lambda_{\phi\Sigma} M_P}{M}   
    \biggr) N_* ~, \\ 
    r &\simeq \frac{12}{N_*^2} + \frac{640}{9} \biggl(\frac{V_\Sigma}{M_P}\biggr)^2 \biggl(
      1 + \frac{2\sqrt{3} \lambda_{\phi\Sigma} M_P }{M}   
      \biggr) ~,
\end{align}
where we explicitly exhibit factors of the Planck mass, which were often set to unity in previous expressions. 
The Starobinsky potential is shown in Fig.~\ref{staro} as the black dotted curve.  The other curves show the deformed potential in Eq.~(\ref{delv}) with $\Delta V$ given by Eq.~(\ref{eq:delvsu5}) for representative values of $V_\Sigma$ and $\lambda_{\phi\Sigma}=0$.

\begin{figure}[ht!]
    \centering
    \includegraphics[width=\textwidth]{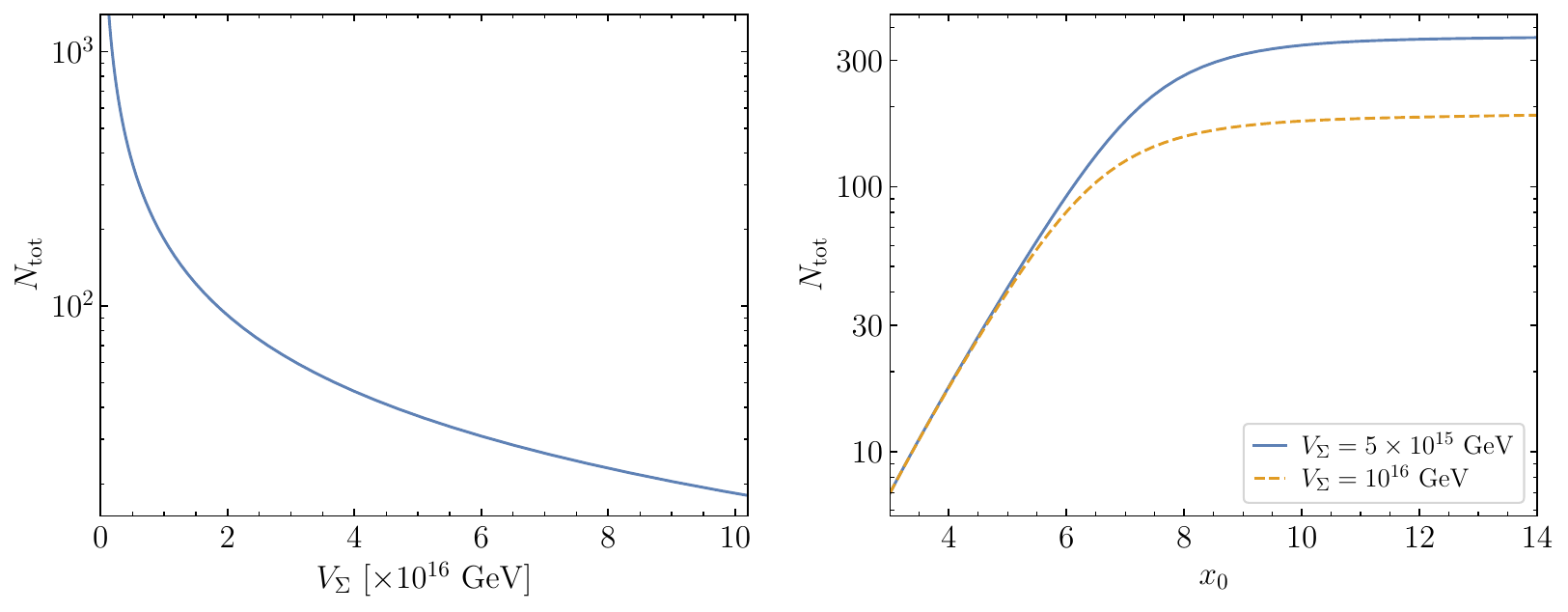}
    \caption{ {\it Left panel:} Total duration of inflation, $N_{\rm tot}$, as a function of $V_{\Sigma}$ with $\lambda_{\phi\Sigma} = 0$. The initial condition is taken as the $x$ value such that $V/M^2=2$. {\it Right panel:} $N_{\rm tot}$, as a function of the inflaton initial condition $x_0$, for two representative values of $V_{\Sigma}$ and $\lambda_{\phi\Sigma} = 0$.}
    \label{staroN}
\end{figure}

We see that the deformed potential deviates noticeably from the Starobinsky model when the inflaton field $x \gtrsim 5$. This feature also appears in Starobinsky-like no-scale supergravity models described by the superpotential (\ref{WZW}) if the trilinear coupling $\lambda_\phi \ne 1$, as was pointed out in~\cite{eno6}. This and related deformations of the no-scale Starobinsky model were discussed recently in~\cite{Antoniadis:2025pfa}. The left panel of Fig.~\ref{staroN} shows the total number of e-folds of inflation, $N_{\rm tot}$ as a function of $V_\Sigma$ assuming an initial value of $x$ corresponding to $V/M^2 = 2$. We see explicitly that $N_{\rm tot}$ diverges as $V_\Sigma \to 0$, corresponding to the Starobinsky case. However, for a given value of $V_\Sigma$, we see in the right panel of Fig.~\ref{staroN} that for an initial inflaton field value, $x_0 \gtrsim 8$, the total number of e-folds becomes constant. That is, the amount of expansion is independent of the initial field value.  This is in sharp contrast to the Starobinsky model where
the total number of e-folds is exponentially sensitive to $x_0$. 

Indeed, one of the criticisms of the Starobinsky model is related to the flatness of the potential at large field values. The possibility that $x$ takes values $x \gg M_P$ threatens the validity of the effective field theory approach (used in most analyses of inflation) and is disfavoured by the Swampland Distance Conjecture \cite{Agrawal:2018own,Lust:2023zql}. The latter implies that a tower of light states appears in the particle spectrum with masses exponentially suppressed by the inflaton field value. It was argued in \cite{ano,Antoniadis:2025pfa} that a deformation similar to that shown in Fig.~\ref{staro} shields the inflaton from taking large field values. Furthermore, inflation with a sufficiently large number of e-folds ($N_{\rm tot} > N_* \sim 50 - 60$) remains largely independent of the initial field value as demonstrated in the right panel of Fig.~\ref{staroN}.  Even for initial values $x > 10$, although the field evolves very quickly initially, slow roll is still possible as friction slows the field dramatically when the plateau at lower field values is reached. The total number of e-folds is largely independent of the initial field value, avoiding the criticism of the undeformed Starobinsky potential that most of the Universe would have evolved from patches with large initial field values. For a more complete discussion of this issue see Ref.~\cite{Antoniadis:2025pfa}.

\begin{figure}[ht!]
  \centering
  \subcaptionbox{\label{nsV} P-ACT-LB }{
  \includegraphics[width=0.48\columnwidth]{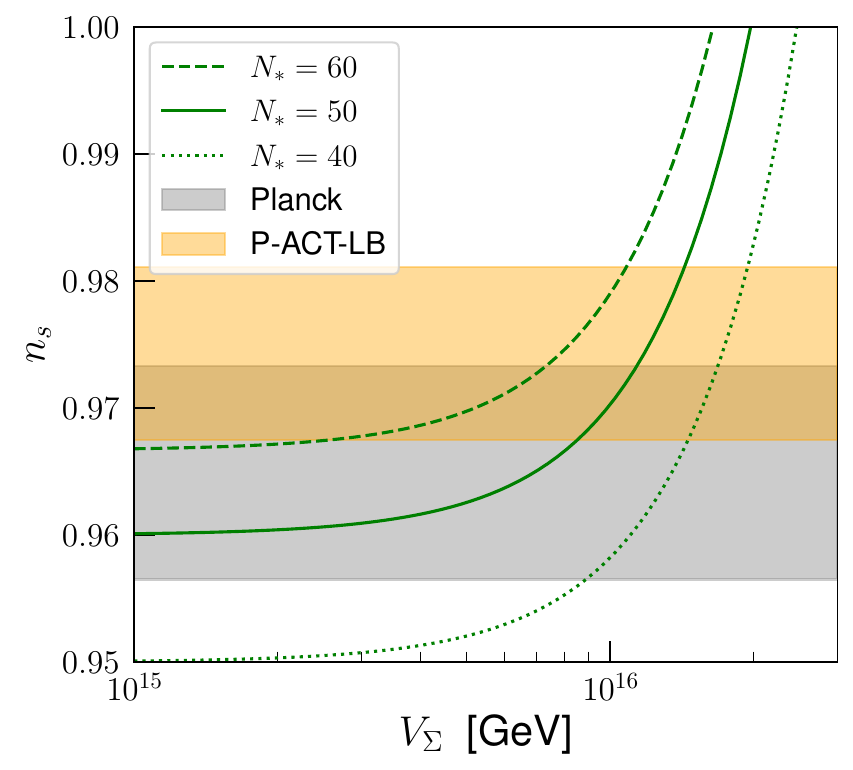}}
  \subcaptionbox{\label{nsV2} P-ACT-SPT}{
  \includegraphics[width=0.48\columnwidth]{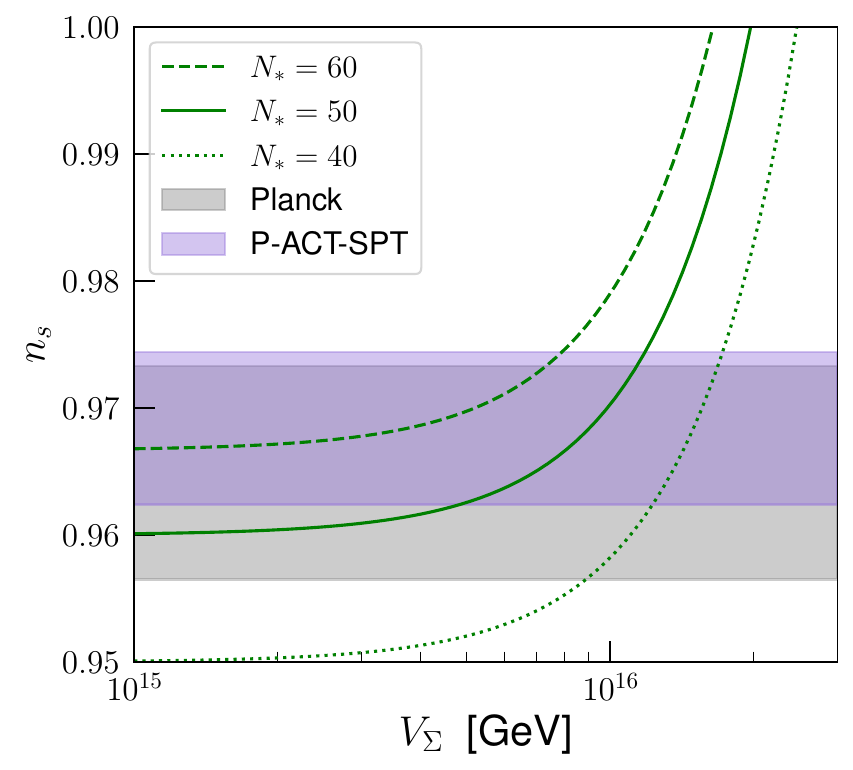}}
\caption{ 
The change in $n_s$ as a function of $V_\Sigma$ for three values of $N_*$, where we set $\lambda_{\phi\Sigma} = 0$, compared with the Planck and P-ACT-LB ranges (left panel) and the Planck and P-ACT-SPT ranges (right panel). } 
  \label{fig:nsandr}
\end{figure}

\begin{figure}[ht!]
    \centering
    \includegraphics[width=.5\textwidth]{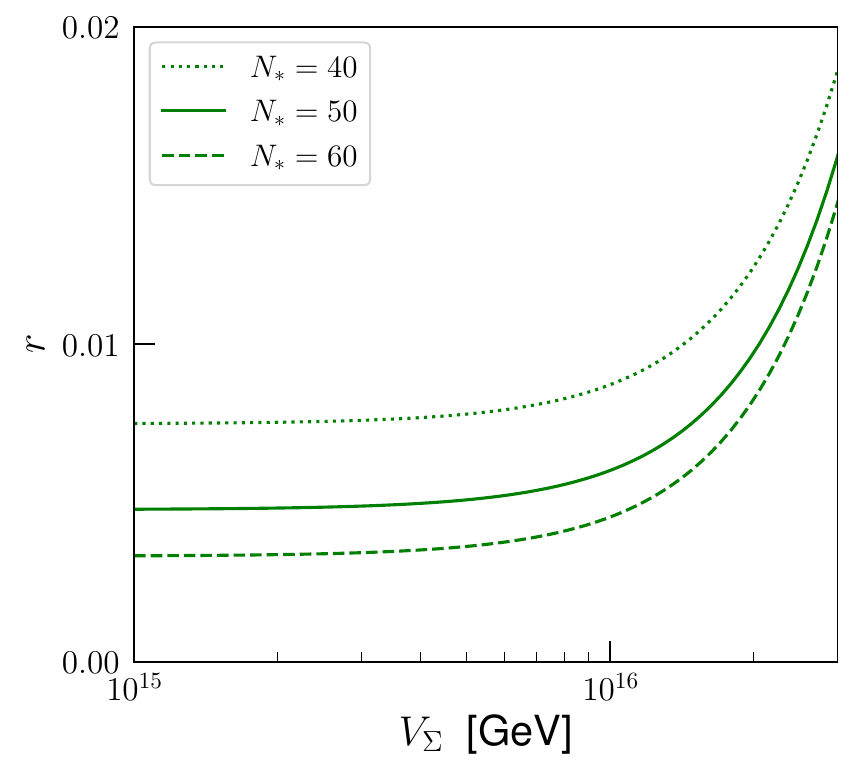}
    \caption{The change in $r$ as a function of $V_\Sigma$ for three values of $N_*$, where we set $\lambda_{\phi\Sigma} = 0$.}
    \label{fig:r}
\end{figure}

We see in Figs.~\ref{fig:nsandr} and \ref{fig:r} that the SU(5) predictions for $n_s$ and $r$ deviate when $V_{\Sigma} \ne 0$ from the Starobinsky values for $V_{\Sigma} = 0$. The gray and yellow shading in Fig.~\ref{nsV} represents the $2\sigma$ Planck range (including lensing)~\cite{Planck} and the $2\sigma$ range from ACT combined with Planck and DESI BAO data (P-ACT-LB)~\cite{ACT:2025fju}. We see that the SU(5) deviation in $n_s$ is constrained by the Planck measurement. For $N_* = 50 \, (60)$, for instance, we have 
\begin{equation}
 \biggl(
    1 + \frac{2 \sqrt{3}\lambda_{\phi\Sigma} M_P }{M}   
    \biggr)^{1/2}   V_\Sigma \, < 1.1 \, (0.73) \times 10^{16}~{\rm GeV}~,
  \label{eq:limonvsigma}
\end{equation}
at the 95\% CL. This estimate shows that the CMB measurement can probe GUT physics directly in this scenario and, indeed, that the Planck measurement has already imposed a severe limit on the GUT scale. Note that this bound exists even if the inflaton has no direct coupling to the GUT Higgs in the superpotential: as we can see in Eq.~\eqref{eq:delvsu5}, $\Delta V \neq 0$ even for $\lambda_{\phi\Sigma} = 0$. This contribution comes from the overall factor of $e^{2K/3}$ in the scalar potential when there are fields that have non-vanishing VEVs during inflation. This type of contribution exists also in the SO(10) model \cite{egnno1} as we discuss further below, but is absent in the flipped SU(5) model \cite{egnno2}.

The constraints on the $(V_\Sigma, n_s)$ plane taking the P-ACT-LB limits on $n_s$ are also shown in Fig.~\ref{nsV}.
In contrast to the Planck measurement, the P-ACT-LB combination of measurements~\cite{ACT:2025fju} prefers the following range of $V_\Sigma$: 
\begin{equation}
 0.85\, (0.26) \times 10^{16}~{\rm GeV} < 
 \biggl(
    1 + \frac{2 \sqrt{3}\lambda_{\phi\Sigma} M_P }{M}   
    \biggr)^{1/2}   V_\Sigma \, < 1.4 \, (1.1) \times 10^{16}~{\rm GeV}~,
  \label{eq:limonvsigma_act}
\end{equation}
for $N_* = 50 \, (60)$. Note the presence of a {\it lower} limit to $V_\Sigma$. It is intriguing that the GUT scale is hinted at by something fundamentally different, namely, observations of inflation.

Similarly, the Planck-ACT-SPT combination for the $(V_\Sigma, n_s)$ plane is shown in Fig.~\ref{nsV2}, and using Eq.~(\ref{CMB-SPA}) for $n_s$ leads to the limits
\begin{equation}
 0.48\, (0) \times 10^{16}~{\rm GeV} < 
 \biggl(
    1 + \frac{2 \sqrt{3}\lambda_{\phi\Sigma} M_P }{M}   
    \biggr)^{1/2}   V_\Sigma \, < 1.2\, (0.79) \times 10^{16}~{\rm GeV}~.
  \label{eq:limonvsigma_spt}
\end{equation}

Fig.~\ref{fig:r} indicates that, whichever of the bounds~\eqref{eq:limonvsigma}, \eqref{eq:limonvsigma_act}, and \eqref{eq:limonvsigma_spt} is considered,  it will be difficult to measure a significant deviation from the Starobinsky value of $r$ in the framework of the SU(5) GUT deformations considered here.

In the present scenario, reheating proceeds through the coupling $\lambda_{H\phi}$ in Eq.~\eqref{W5}, through which the inflaton decays into $H_u$ and $H_d$ at a rate given by 
\begin{equation}\label{eq:gamma5}
  \Gamma (\phi \to H_u H_d) \simeq 2 \times \frac{|\lambda_{H\phi}|^2}{8 \pi} M ~.
\end{equation}
In order to evade the gravitino overproduction problem \cite{Ellis:2015jpg}, one must impose $|\lambda_{H\phi}| \lesssim 10^{-5}$. If we also introduce right-handed neutrinos to this model, which are SU(5) singlets with $R$-parity odd, we can couple the inflaton also to these fields without modifying the inflation dynamics. In this case, the inflaton can decay into right-handed neutrinos as well, and the gravitino overproduction bound again restricts the inflaton-right-handed neutrino couplings to be $\lesssim 10^{-5}$. 

The temperature at reheating, $T_{\rm RH}$, is often defined by the condition
$\rho_r(a_{\rm RH}) = \rho_\phi(a_{\rm RH})$, or~
\beq
	\frac{g_{\rm RH} \pi^2}{30} T_{\rm RH}^4 = \frac{12}{25} \left(\Gamma_\phi M_P \right)^2 \, .
 \label{deftrh}
\eeq
Therefore, for $M\simeq 3 \times 10^{13}$~GeV we have 
\beq
T_{\rm RH} \simeq 6.7 \times 10^{14}\ |\lambda_{H\phi} |~{\rm GeV}  \, , 
\eeq
where we have used $g_{\rm RH} = 915/4$.

\begin{figure}[ht!]
\centering
\includegraphics[width=0.60\columnwidth]{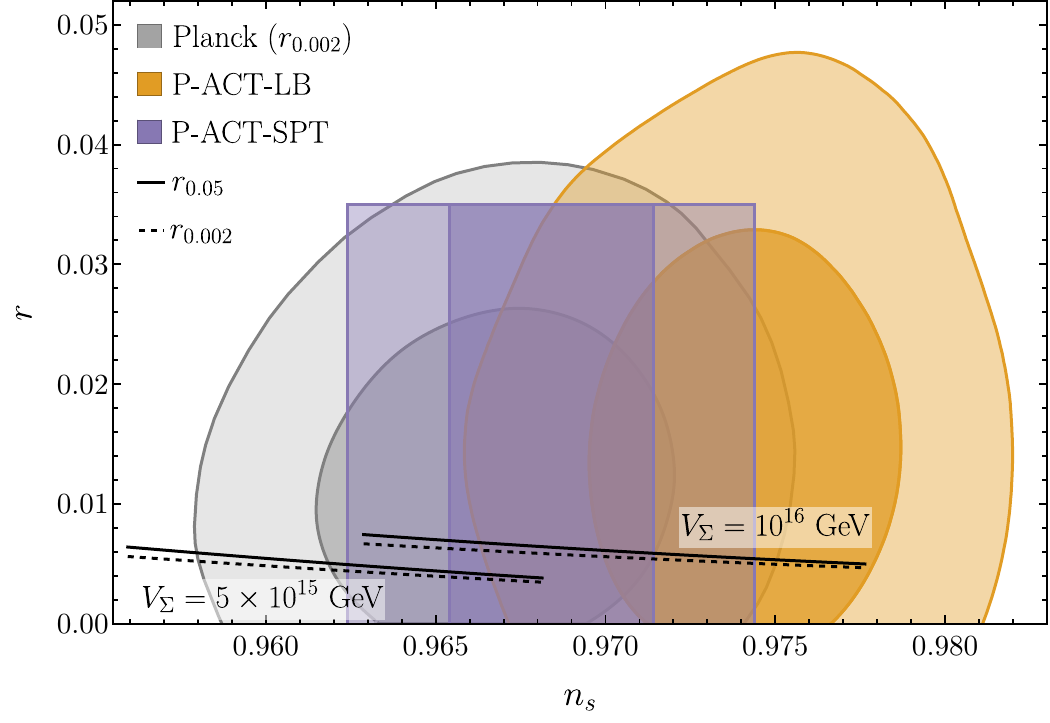}
\caption{ CMB observables for the minimal SU(5) model for two values of $V_{\Sigma}$ with $\lambda_{\phi\Sigma}=0$. Shown as the shaded regions are the $1\sigma$ and $2\sigma$ contours for Planck (gray), the P-ACT-LB results (orange), and the P-ACT-SPT results for $n_s$ only (purple). Each curve covers the allowed phenomenological range for $N_*$, with the continuous curves corresponding to the evaluation of $r$ at the Planck/ACT pivot scale $k_*=0.05\,{\rm Mpc}^{-1}$, while the dashed line corresponds an evaluation of $r$ at the WMAP pivot scale $k_*=0.002\,{\rm Mpc}^{-1}$ (see Fig.~\ref{fig:Nstar5}).}
  \label{fig:cmbsu5}
\end{figure}

As is commonly done, we can also visualize the CMB observables for the SU(5) model in the $(n_s,r)$ plane, as shown in Fig.~\ref{fig:cmbsu5}. The $1\sigma$ and $2\sigma$ confidence level contours for the Planck (+BICEP/Keck) analysis are shown in grey. For this dataset the constraints on the tensor-to-scalar ratio are provided at the WMAP pivot scale $k_*=0.002\ {\rm Mpc}^{-1}$~\cite{rlimit}. The P-ACT-LB results are shown in orange. For these the Planck pivot scale ($k_* = 0.05\ {\rm Mpc}^{-1}$) is used for both inflationary observables~\cite{ACT:2025fju,ACT:2025tim}. The P-ACT-SPT constraint on $n_s$ (only) is also included, in purple. The curves shown in the Figure correspond to the evaluation of the inflationary observables for two values of $V_{\Sigma}$ with $\lambda_{\phi\Sigma}=0$, the solid ones at the Planck/ACT pivot scale for $r$, and the dashed ones at the WMAP scale. As one can see, there is very little difference between the two choices of $k_*$. The curves span the allowed range of couplings $\lambda_{H\phi}$ for which the decay rate (\ref{eq:gamma5}) leads to reheating temperatures above the Big Bang Nucleosynthesis scale, $T_{\rm RH}\gtrsim 4\ {\rm MeV}$~\cite{tr4}, while maintaining perturbativity, conservatively taken as $\lambda_{H\phi}<1$. This corresponds to $41\lesssim N_*\lesssim 55$, as can be seen from (\ref{eq:nstarreh}) together with~\cite{egnov}
\beq\label{eq:nstargamma}
N_* \;\supset\; \frac{1-3 w_{\mathrm{int}}}{12\left(1+w_{\mathrm{int}}\right)} \ln \left(\frac{\rho_{\mathrm{R}}}{\rho_{\text {end }}}\right) \;\simeq\; \frac{1}{6}\ln\left(\frac{\Gamma}{H_{\rm end}}\right)\,.
\eeq
As can be seen in Fig.~\ref{fig:nsandr}, a larger (yet typical) value of  $V_\Sigma = 2 \times 10^{16}$~GeV is in tension with all of the observations. A curve with this value of $V_\Sigma$ would lie outside the range plotted in Fig.~\ref{fig:cmbsu5}.


\begin{figure}[ht!]
\centering
\includegraphics[width=0.49\columnwidth]{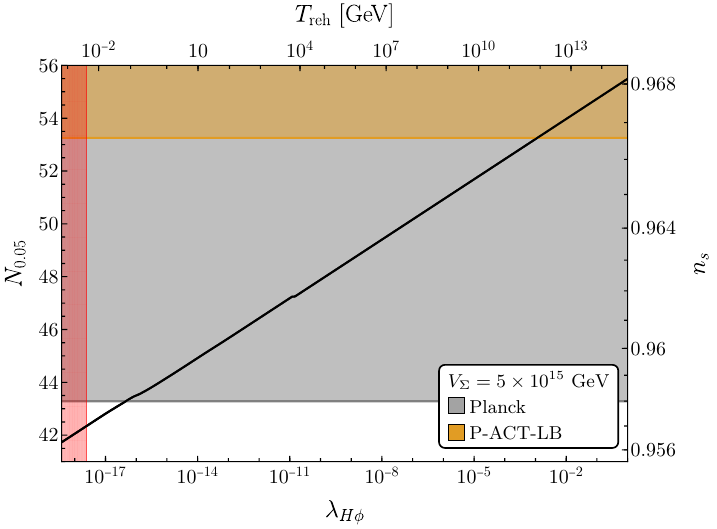} \includegraphics[width=0.49\columnwidth]{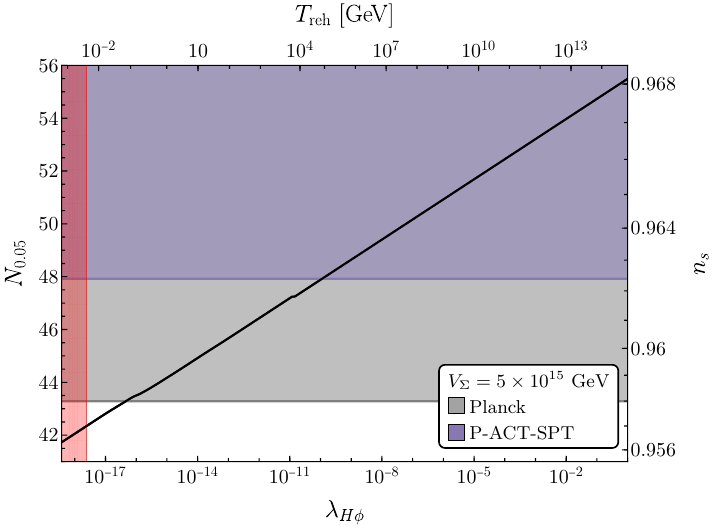}\\
\includegraphics[width=0.49\columnwidth]{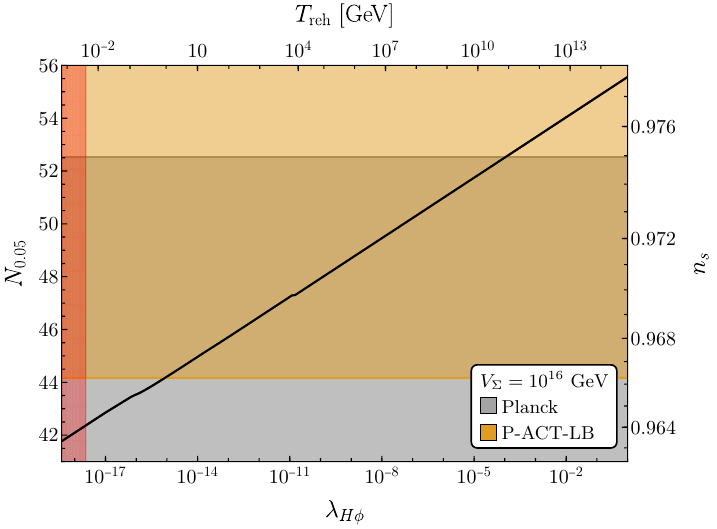} \includegraphics[width=0.49\columnwidth]{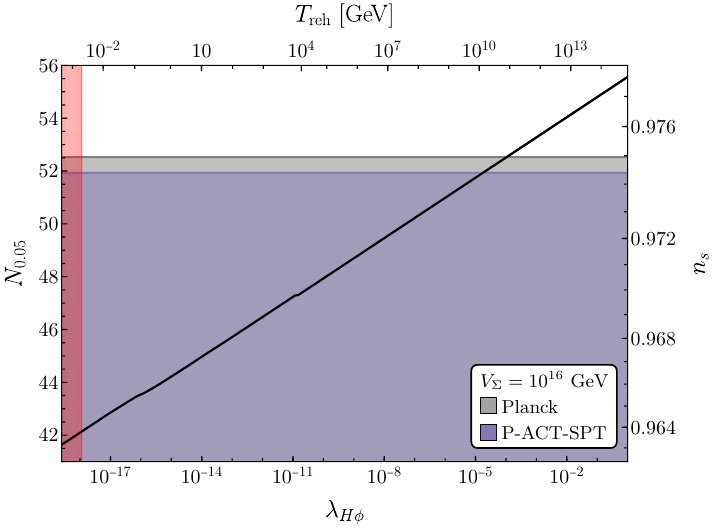}
\caption{ The range of $N_*$ (and $n_s$) at the Planck/ACT pivot scale for the SU(5) model with the two values of $V_{\Sigma}$ considered in Fig.~\ref{fig:cmbsu5}, as functions of $\lambda_{H\phi}$ (or equivalently $T_{\rm reh}$). The red vertical shaded region is forbidden by BBN, while the horizontal shaded regions correspond to the $2\sigma$ CL values for $n_s$ from Planck, P-ACT-LB (left) and P-ACT-SPT (right). }
  \label{fig:Nstar5}
\end{figure}

More explicitly, the dependence of the number of e-folds at the Planck/ACT pivot scale on the coupling $\lambda_{H\phi}$ can be seen in Fig.~\ref{fig:Nstar5}. Each row corresponds to the labeled value of $V_{\Sigma}$ considered in Fig.~\ref{fig:cmbsu5} above, which uniquely determines the dependence of $N_*$ on $n_s$ via $V_*$ in (\ref{eq:nstarreh}). In the left panels, the shading corresponds to the 2$\sigma$ limits on $n_s$ taken from Fig.~\ref{fig:cmbsu5} using Planck (grey) and P-ACT-LB (orange).
Values of $N_* \sim 56$ are close to the limit of instantaneous reheating. That is, no values of $\lambda_{H\phi}$ will yield larger values of $N_*$. The black curves follow the relation $N_*\sim \frac{1}{3}\ln|\lambda_{H\phi}|$ except for small `breaks' in the line which signal significant changes in the number of degrees of freedom $g_{\rm RH}$. 
The gravitino bound of $\lambda_{H\phi} \lesssim 10^{-5}$ cannot be satisfied simultaneously with the P-ACT-LB data set for $V_\Sigma = 5 \times 10^{15}$~GeV, but can be for $V_\Sigma = 10^{16}$~GeV. In the right panels, we show the  $2\sigma$ limits (from Fig.~\ref{fig:cmbsu5}) from Planck (grey)
and P-ACT-SPT (purple). The BBN limit is shown as the forbidden region in red on the left side of each panel. 


The bounds~\eqref{eq:limonvsigma}, \eqref{eq:limonvsigma_act}, and \eqref{eq:limonvsigma_spt} have important implications on the type of supersymmetric SU(5) models allowed. The value of $V_\Sigma$ can be determined by a renormalization-group analysis. Using threshold corrections at the GUT scale, we can determine a combination of the masses of the GUT-scale particles, namely $(M_X^2 M_\Sigma)^{1/3}$, where $M_\Sigma$ is the mass of the adjoint Higgs field, and the VEV of the GUT Higgs is given by \cite{Ellis:2016tjc}
\begin{equation}
  V_{\Sigma} = \frac{1}{5} \biggl(\frac{2}{\lambda^\prime g_5^2}\biggr)^{1/3} 
  (M_X^2 M_\Sigma)^{1/3}  = \frac{1}{5 \lambda} M_{H_C} ~.
\end{equation}
The SU(5) gauge coupling, $g_5$, is also determined by GUT threshold conditions. As a result, $V_\Sigma$ is given as a function of the undetermined parameter $\lambda^\prime$ in the SU(5) superpotential (\ref{W5}), once the MSSM mass spectrum is fixed. The bound~\eqref{eq:limonvsigma} then leads to a lower limit on $\lambda^\prime$, which provides an additional restriction on the parameter space. 

In minimal SU(5) with supersymmetry-breaking boundary conditions similar to those in the constrained minimal supersymmetric Standard Model (CMSSM), the bound in Eq.~(\ref{eq:limonvsigma}) is hard to satisfy. In \cite{Ellis:2020mno}, we considered a variety of models in the context of no-scale SU(5). These models were constructed to obtain the correct relic density, Higgs mass, and an acceptable proton lifetime. For the latter, it was generally found that $\lambda'$ should be small so as to obtain a large mass for $H_C$ in order to suppress the dimension-5 proton decay operators. As a consequence, since $M_{H_C} = 5 \lambda V_\Sigma$, 
we require a large value of $V_\Sigma$ in excess of the bound in Eq.~(\ref{eq:limonvsigma}).
In this class of models, there are six input parameters in addition to the sign of the $\mu$-term. These are the gaugino mass, gravitino mass, universality input scale, the two GUT Higgs couplings, and the ratio of MSSM Higgs vacuum expectation values:~\footnote{Treating both $\lambda$ and $\lambda'$ as free parameters requires the inclusion of the dimension-5 operator, $ W \supset \frac{c_5}{M_P} {\rm Tr}\left[
\Sigma {\cal W} {\cal W}
\right] $, where ${\cal W}\equiv  {\cal W}^A T^A$ denotes the
superfields corresponding to the field strengths of the SU(5) gauge vector bosons. The inclusion of these terms is important when considering gauge coupling unification. See \cite{Ellis:2020mno} for more details.}
\beq
 m_{1/2},\ m_{3/2},\ M_{in},\ \lambda,\ \lambda', \ \tan\beta, \ \rm{sign}(\mu).
\eeq
As an example, we consider the model M1 of \cite{Ellis:2020mno}. In this (no-scale supergravity) model, universality boundary conditions are set at $M_{\rm in} = 10^{16.5}$~GeV. At that scale the gaugino mass $M_5 = m_{1/2}$. This is run down to the GUT scale (defined when the electroweak gauge couplings are equal, $g_1 = g_2$). The sfermions and the Higgs adjoint are taken to be untwisted fields. As a consequence, their soft scalar masses $m_{\bf 10} = m_{\bf {\bar 5}} = m_\Sigma = 0 $  and hence are also universal at $M_{\rm in}$. In contrast, 
the two Higgs five-plets are taken to be twisted fields with masses $m_H = m_{\overline{H}} = m_{3/2}$. The trilinear supersymmetry breaking terms are $A_{\bf 10} = A_{\overline{\bf 5}} = m_{3/2}$, $A_\lambda = 2 m_{3/2}$, and $A_{\lambda'} = 0$. Similarly, the bilinear terms are $B_H = 2 m_{3/2}$ and $B_\Sigma = 0$. For further details see \cite{Ellis:2020mno}. 

As in M1 of \cite{Ellis:2020mno}, we set $m_{1/2} = 6000$~GeV, $\tan \beta = 6$, and $\lambda = 0.6$. For $\lambda' = 10^{-5}$,
taking $m_{3/2} = 9070$~GeV we obtain a Higgsino-like LSP with mass $\simeq 1.1$~TeV
providing a relic density $\Omega_{\tilde H} h^2 \simeq 0.12$. For this choice of $m_{1/2}$, the calculated Higgs mass using {\tt FeynHiggs} is 124.3 GeV, which is consistent with the experimental value within the theoretical uncertainties. The dominant contribution to the proton lifetime stems from dimension-5 operators and was found to be $\tau_p (p \to K^+ {\bar \nu})  \simeq 9 \times 10^{33}$ years, which is compatible with the lower bound from experiment of $6.6 \times 10^{33}$ years. However, for $\lambda' = 10^{-5}$, the colored Higgs mass is $\simeq 3.5 \times 10^{17}$~GeV, and $V_\Sigma \simeq 1.2 \times 10^{17}$~GeV, an order of magnitude in excess of the bound in Eqs.~(\ref{eq:limonvsigma})--(\ref{eq:limonvsigma_spt}).

\begin{figure}[ht!]
  \centering
  \subcaptionbox{\label{fig:cmssm_v} $V_\Sigma$ }{
  \includegraphics[width=0.5\columnwidth]{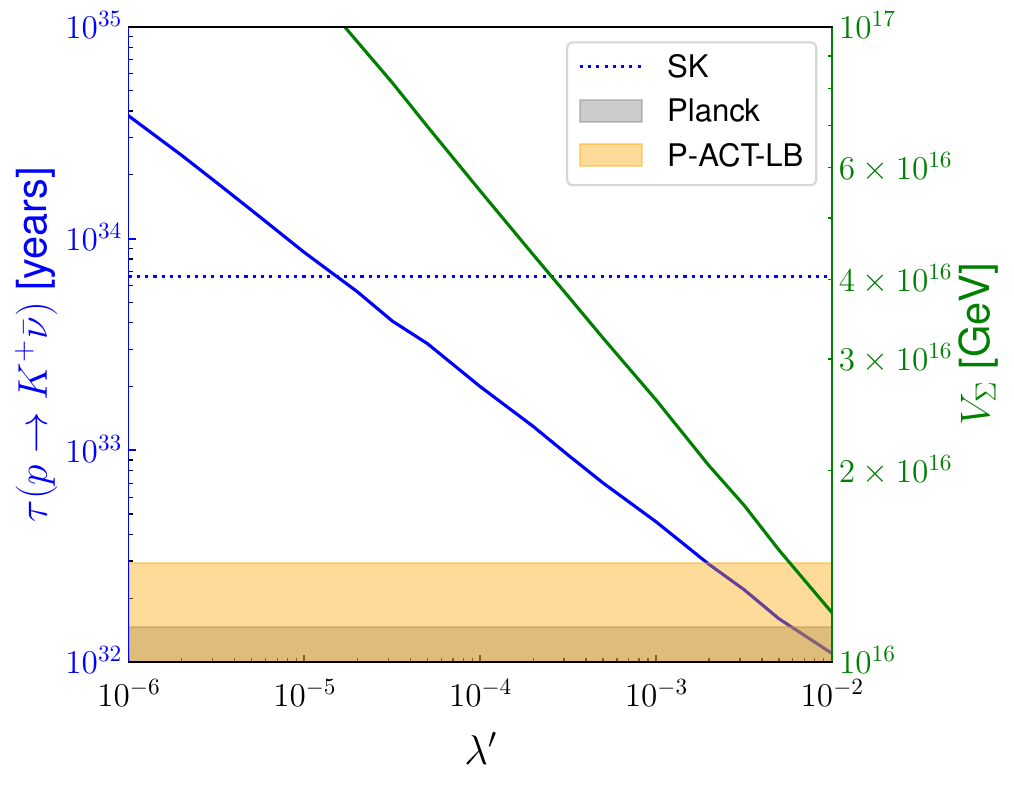}}
  \subcaptionbox{\label{fig:cmssm_ns} $n_s$}{
  \includegraphics[width=0.47\columnwidth]{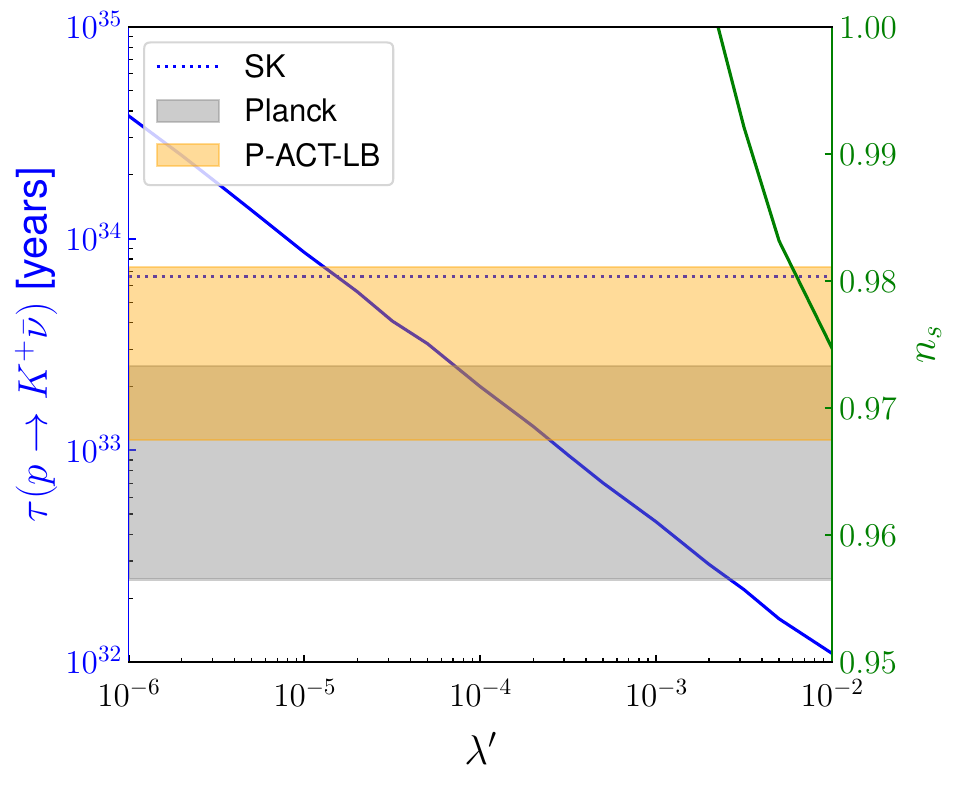}}
\caption{ (a)
{\it Left panel:} Values of the expectation value $V_\Sigma$ (green, right axis) and proton decay lifetime for $p \to K^+ {\bar \nu}$ (blue, left axis). The blue dotted line corresponds to the lower limit on the proton lifetime, and the gray (orange) area corresponds to the limit on $V_\Sigma $ obtained from Planck data, Eq.~\eqref{eq:limonvsigma} (from the P-ACT-LB combination of measurements, Eq.~\eqref{eq:limonvsigma_act}) for $N_* = 50$. 
(b) {\it Right panel:} As in the left panel, but the green line and axis show values of $n_s$. 
} 
  \label{M1}
\end{figure}

The value of $V_\Sigma$ is directly related to the choice of $\lambda'$. In Fig.~\ref{M1},
we allow $\lambda'$ to vary holding the other inputs fixed, except $m_{3/2}$.  For each choice of $\lambda'$, $m_{3/2}$ is adjusted so that the relic density is fixed at $\Omega h^2 = 0.12$. Over the range in $\lambda' = 10^{-6} - 10^{-2}$ displayed in Fig.~\ref{M1}, $m_{3/2}$ varies from 8.7--10.2~TeV, and the lightest MSSM Higgs mass varies from 124.2--124.7~GeV. Fig.~\ref{fig:cmssm_v} shows the calculated values of $V_\Sigma$ (green curve, right axis) and $\tau_p$ (blue curve, left axis) as a function of $\lambda'$. The horizontal blue dotted line shows the lower limit to $\tau (p \to K^+ \bar{\nu})$, $\tau (p \to K^+ \bar{\nu}) > 6.6\times 10^{33}$~years \cite{Super-Kamiokande:2014otb, Takhistov:2016eqm}, and the gray (orange) area shows the limit on $V_\Sigma $ from Eq.~\eqref{eq:limonvsigma} (Eq.~\eqref{eq:limonvsigma_act}) for $N_* = 50$. As one can see, the value of $V_\Sigma$ decreases significantly as $\lambda'$ is increased over the displayed range, and satisfies the bound in Eq.~(\ref{eq:limonvsigma}) only when $\lambda' \gtrsim 0.01$ for the Planck data. The limit is slightly relaxed for the higher value of $V_\Sigma$ allowed by the P-ACT-LB combination of data.  However, the proton lifetime also decreases significantly, and $\tau_p \lesssim 1.1 \times 10^{32}$ years for $\lambda' \gtrsim 0.01$.
Finally, we note that playing on the uncertainty in the calculated value of the MSSM Higgs boson, $m_h$,
we can reduce $V_\Sigma$ to within a factor of 2 of the upper bound, at the expense of taking the universal gaugino mass to 20 TeV, resulting in a Higgs mass of $127.4 \pm 0.7$~GeV. Clearly this is not an ideal solution.

Another way to visualize the problem is displayed in Fig.~\ref{fig:cmssm_ns} which shows the calculated value of $n_s$ as the green curve corresponding to the right axis (the blue curve is the same as in panel a). The Planck upper limit to $n_s$ requires $\lambda' > 10^{-2}$ while the proton life-time limit requires $\lambda' \lesssim 2 \times 10^{-5}$. The ACT upper limit to $n_s$ only relaxes this limit to $\lambda' \gtrsim .006$.

\begin{figure}[ht!]
  \centering
  \subcaptionbox{\label{fig:cmssm_v_spt} $V_\Sigma$ }{
  \includegraphics[width=0.5\columnwidth]{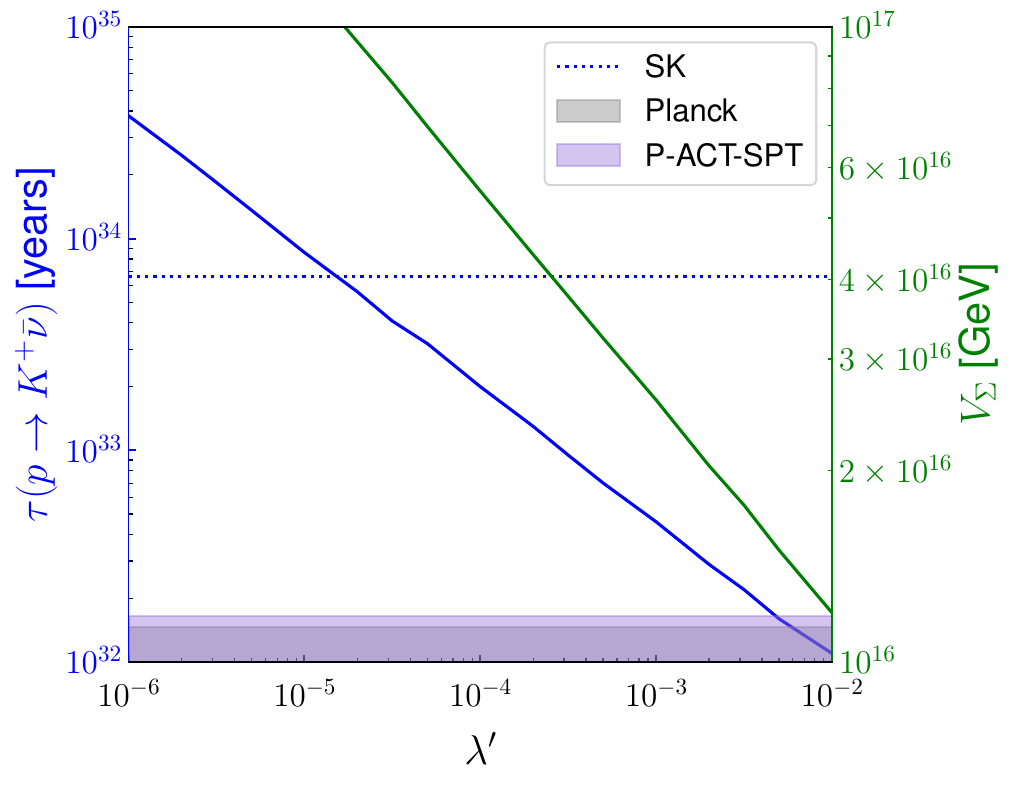}}
  \subcaptionbox{\label{fig:cmssm_ns_spt} $n_s$}{
  \includegraphics[width=0.47\columnwidth]{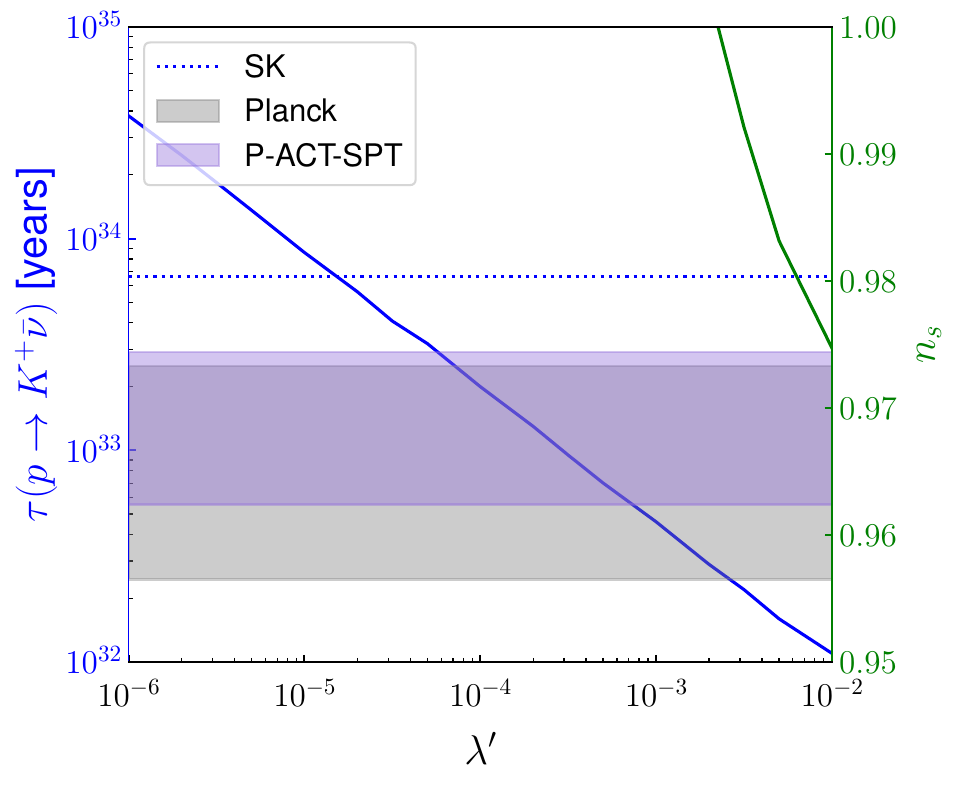}}
\caption{ As in Fig.~\ref{M1}, but showing the constraints from the P-ACT-SPT data combination.
} 
  \label{fig:cmssm_spt}
\end{figure}

The limits coming from the P-ACT-SPT data combination are shown in Fig.~\ref{fig:cmssm_spt}.
Since the upper limit on $n_s$ is tighter (and very close to the limit from Planck alone), the lower limit to $\lambda' \gtrsim 0.01$ still remains. 

As one can see from Eq.~(\ref{eq:limonvsigma}), compatibility between the limits from proton decay and the CMB spectrum can be achieved at the expense of fine-tuning $\lambda_{\phi\Sigma}$. For $\lambda_{\phi\Sigma} \simeq -M/2\sqrt{3}M_P$, $V_\Sigma$ can  be made large enough to satisfy the proton decay bound and yet still satisfy the upper bound in Eq.~(\ref{eq:limonvsigma}). For example, for $V_\Sigma = 1.2 \times 10^{17}$~GeV, we require
$\lambda_{\phi\Sigma} = (-0.345 \pm 0.002) \times 10^{-5}$ for $N_* = 50$ to obtain a value of $n_s$ inside the 2$\sigma$ range determined by Planck. Similar fine-tunings would be necessary for the other combinations of CMB data or other values of $N_*$.

As an alternative, in \cite{Evans:2019oyw}, a minimal SU(5) supersymmetric GUT model was considered in the context of pure gravity mediation (PGM)~\cite{pgm,eioy,eioy2, Evans:2014xpa,eioy5}, a variant of split supersymmetry \cite{split}. In this model, all fields other than $T$ and $\phi$ (the inflation sector) are taken as twisted so that their scalar masses are set by the gravitino mass and near the PeV scale.  In contrast, the gaugino masses are fixed at $M_{\rm in}$ by anomaly mediation \cite{anom,ggw,mAMSB}:
\begin{equation}
 M_5 = \frac{b_5 g_5^2}{16\pi^2} m_{3/2} ~,
\label{eq:m5anom}
\end{equation}
with $b_5 = -3$ the $\beta$-function coefficient of the SU(5) gauge
coupling. Thus, above the GUT scale the gauginos acquire a universal mass
that is orders of magnitude smaller than the scalar masses. 
As a consequence the LSP is typically wino-like.~\footnote{It is also possible to construct PGM models with a Higgsino-like LSP \cite{eioy5,Evans:2022gom}, but we do not discuss them here.} 
The trilinear soft supersymmetry-breaking terms $A_i$ are also set by anomaly mediation at $M_{\rm in}$ and these can be treated as vanishing. The bilinear supersymmetry-breaking terms $B_H$ and $B_\Sigma$ are both set by the supergravity relation $B_i = A_i - m_{3/2}$ to $-m_{3/2}$ at $M_{\rm in}$. Relative to the previous model,
there is one fewer free parameter so that our inputs are now
\beq
 m_{3/2},\ M_{in},\ \lambda,\ \lambda', \ \tan\beta, \ \rm{sign}(\mu).
\eeq
In this case, for relatively low $\tan \beta \simeq 3$ and a large (PeV-scale) value for $m_{3/2}$, the correct relic density can be obtained with a Higgs mass $m_h \simeq 125$~GeV.

Due to the significantly larger scalar masses, the dimension-5 operators leading to proton decay are suppressed \cite{Hisano:2013exa, McKeen:2013dma, Nagata:2013sba, Nagata:2013ive,
Evans:2015bxa, eelnos}. Indeed, it was shown in Ref.~\cite{Evans:2015bxa} that in
minimal supersymmetric SU(5) with PGM, the lifetime of the $p \to K^+
\bar{\nu}$ mode is much longer than the current experimental bound for
$m_{3/2} \gtrsim 100$~TeV. 
When dimension-five proton decay is suppressed,  dimension-six
proton decay induced by the exchange of GUT gauge bosons becomes
dominant. The primary decay channel is $p \to  \pi^0 e^+$ and its rate goes as
$\propto M_X^{-4}$ where the $X$ gauge boson mass is given by $5 g_5 V_\Sigma$. The predicted lifetime is typically above the current
experimental bound, $\tau (p \to \pi^0 e^+) > 1.6\times 10^{34}$~years
\cite{Miura:2016krn}. 

\begin{figure}[ht!]
  \centering
  \subcaptionbox{\label{fig:pgm_v} $V_\Sigma$ }{
  \includegraphics[width=0.48\columnwidth]{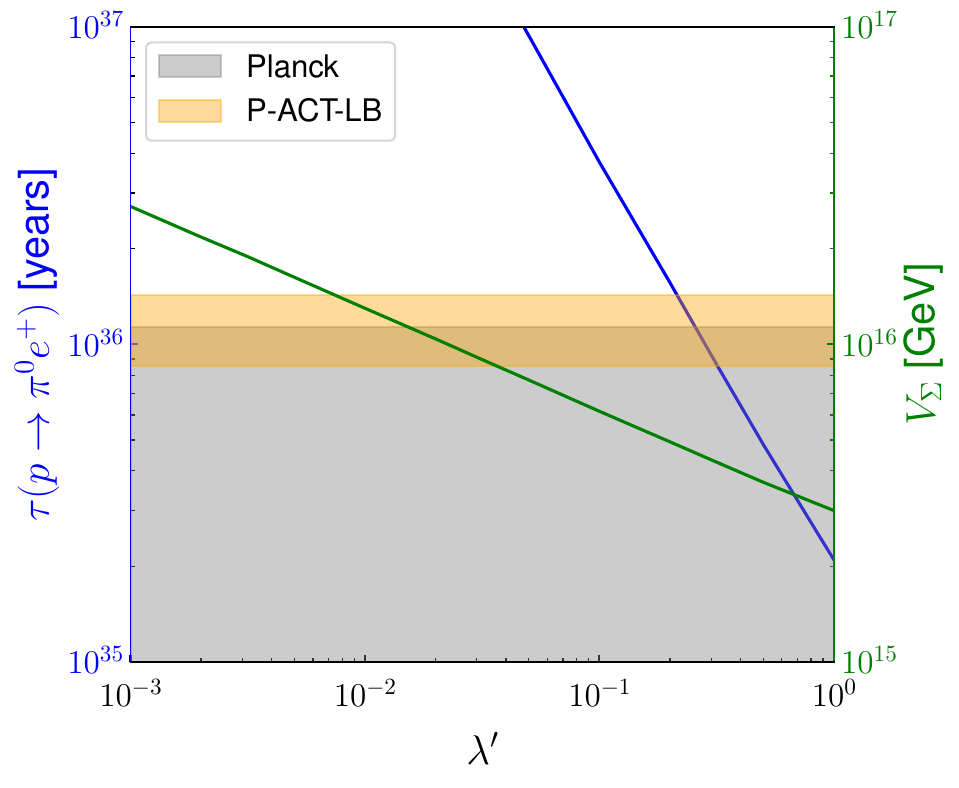}}
  \subcaptionbox{\label{fig:pgm_ns} $n_s$}{
  \includegraphics[width=0.48\columnwidth]{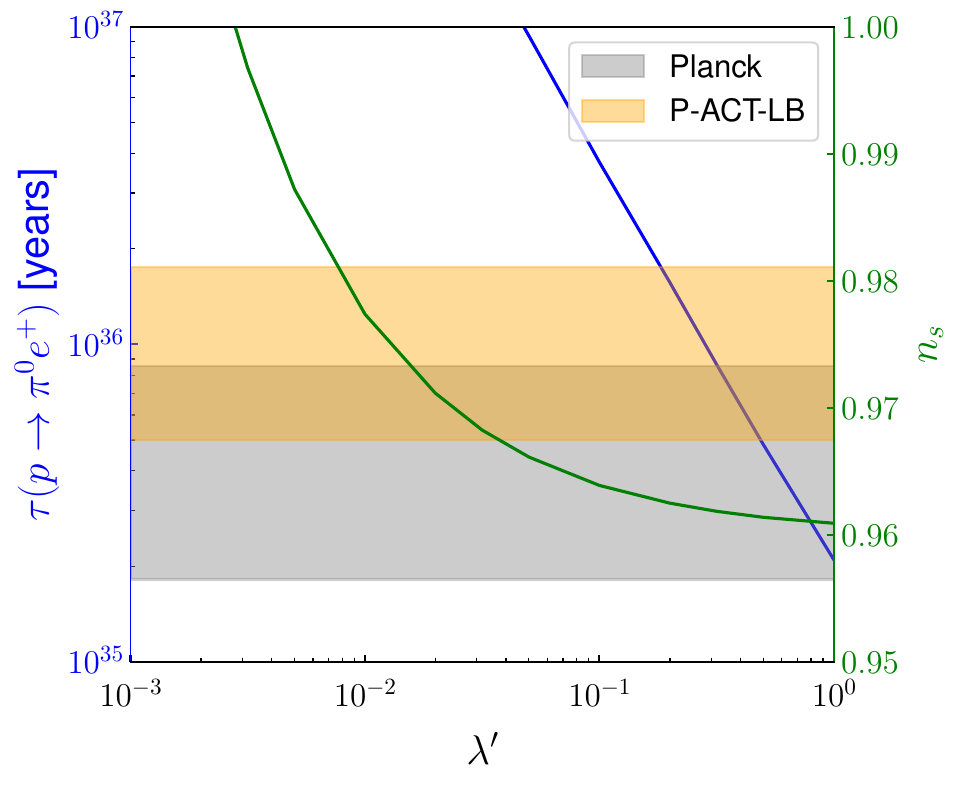}}
\caption{ (a) {\it Left panel:}
The expectation value, $V_\Sigma$ (green, right axis) and lifetime for $p \to \pi^0 e^+$ decay (blue, left axis). The gray (orange) area corresponds to the limit on $V_\Sigma$ from Eq.~\eqref{eq:limonvsigma} (Eq.~\eqref{eq:limonvsigma_act}) for $N_* = 50$. 
(b) {\it Right panel:} As in the left panel, but the green line and axis show values of $n_s$. 
} 
  \label{fig:pgm_50}
\end{figure}

\begin{figure}[ht!]
  \centering
  \subcaptionbox{\label{fig:pgm_v_spt} $V_\Sigma$ }{
  \includegraphics[width=0.48\columnwidth]{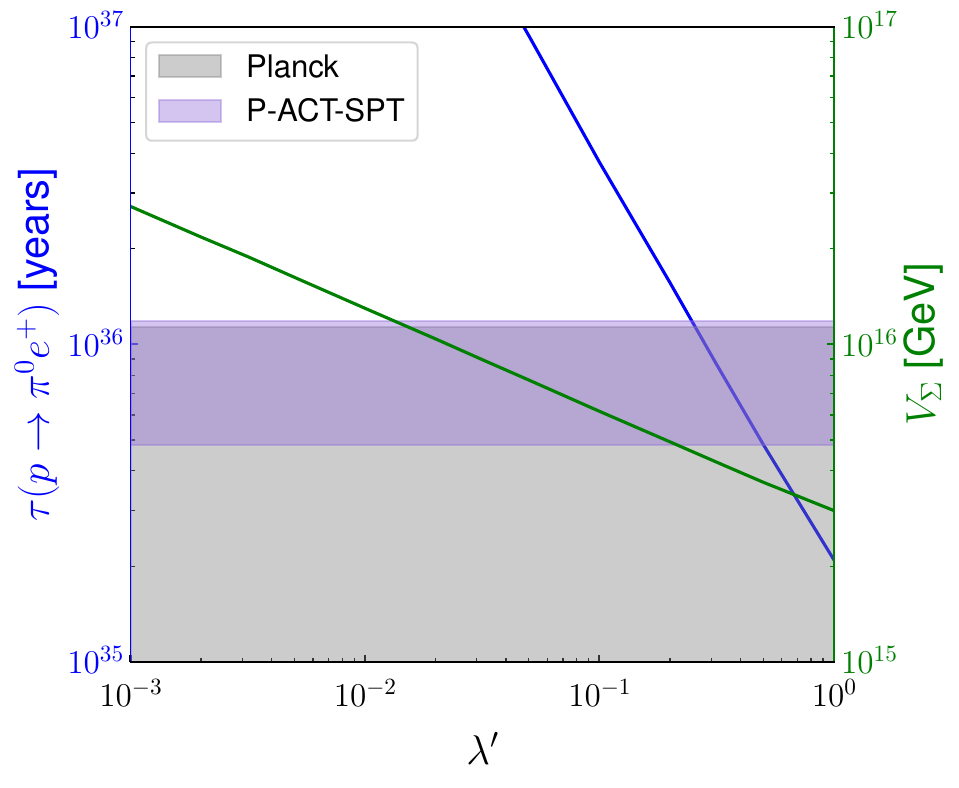}}
  \subcaptionbox{\label{fig:pgm_ns_spt} $n_s$}{
  \includegraphics[width=0.48\columnwidth]{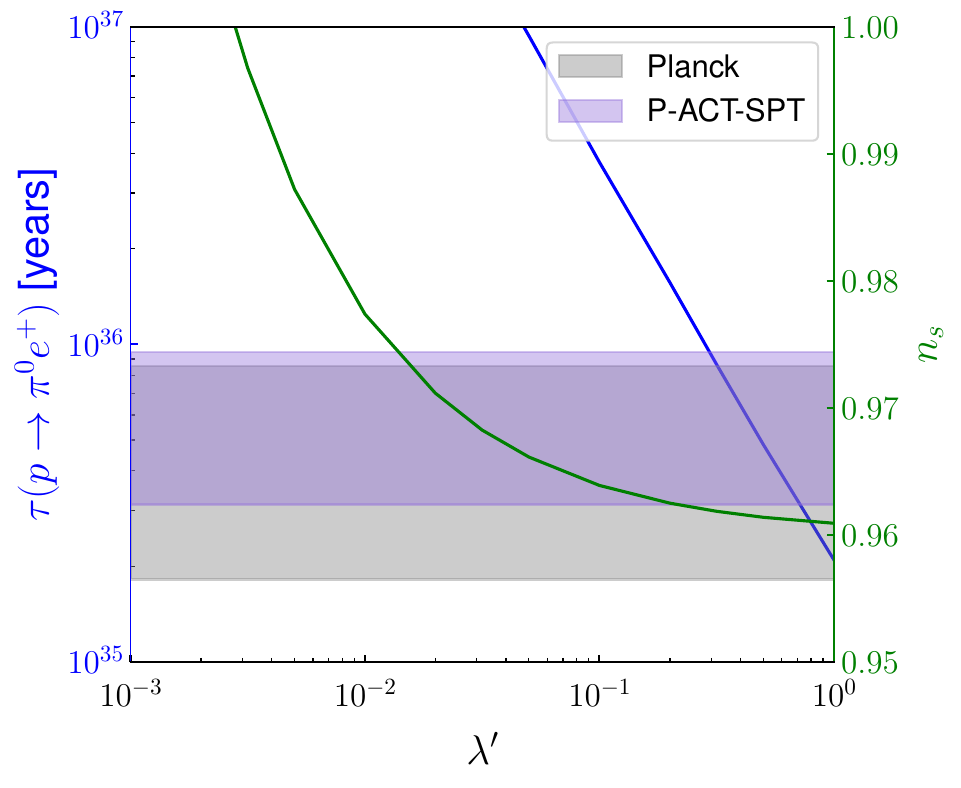}}
\caption{As in Fig.~\ref{fig:pgm_50}, but showing the constraints from the P-ACT-SPT data combination for $N_* = 50$.  } 
  \label{fig:pgm_50_spt}
\end{figure}

In the left panel of Fig.~\ref{fig:pgm_50} we plot $V_\Sigma$ (green, right axis) and $\tau_p$~($p \to \pi^0 e^+$) (blue, left axis) as functions of $\lambda'$. Here we have chosen $M_{\rm in} = 10^{18}$~GeV, $\lambda = 1$, $\tan \beta = 3$,  $\mu < 0$ and $N_* = 50$. From Eq.~(\ref{eq:limonvsigma}) based on Planck data alone, we have the restriction that $\lambda' > 0.016$ for $N_* = 50$. Including data from ACT and BAO (Eq.~(\ref{eq:limonvsigma_act})), the limit is relaxed to $\lambda' > 0.008$. 
Once again, the P-ACT-SPT data combination gives a lower bound to $\lambda'$ similar to that of Planck alone as seen in Fig.~\ref{fig:pgm_50_spt}. The gravitino mass is adjusted to obtain the correct relic density
and varies from 260--575~TeV for $\lambda' = 1 - 10^{-3}$. For these choices, the wino mass is about 3.1 TeV and the lightest MSSM Higgs mass ranges between 123.8--125.6 GeV. The shortest proton lifetime is $\tau_p = 2.1 \times 10^{35}$ years at $\lambda' = 1$. The right panel of Fig.~\ref{fig:pgm_50} is similar to the left panel, but with values of $n_s$ shown in green. We see that there are ranges of $\lambda'$ that are compatible with either the data from Planck and/or the P-ACT-LB combination and/or the P-ACT-SPT combination.

\begin{figure}
  \centering
  \subcaptionbox{\label{fig:pgm50} $N_* = 50$ }{
  \includegraphics[width=0.48\columnwidth]{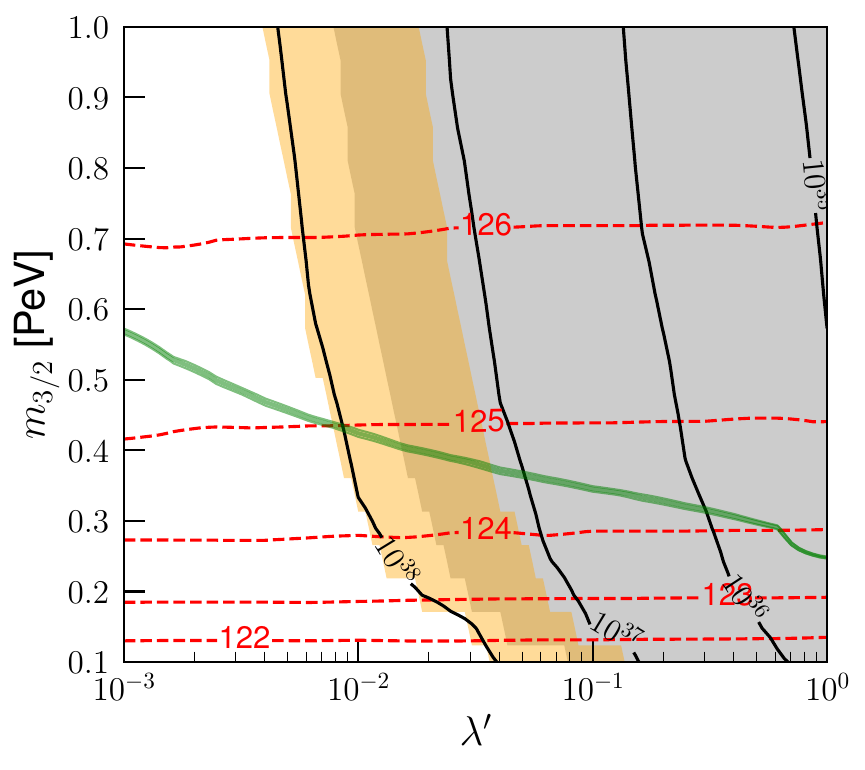}}
  \subcaptionbox{\label{fig:pgm60} $N_* = 60$}{
  \includegraphics[width=0.48\columnwidth]{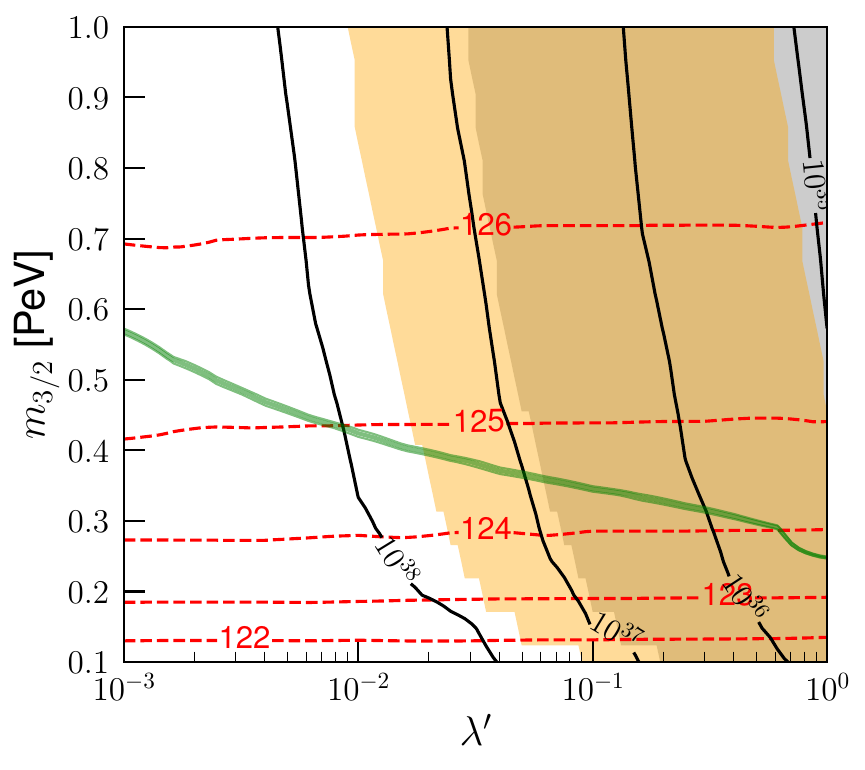}}
\caption{ The $(\lambda',m_{3/2})$ planes in minimal supersymmetric SU(5) with PGM for $N_* = 50$ {\it (left panel)} and $N_* = 60$ {\it (right panel)}, both with fixed $M_{\rm in} = 10^{18}$~GeV, $\lambda = 1$, $\tan \beta = 3$ and $\mu < 0$. Red dashed lines denote the Higgs mass [GeV], the black solid lines show $\tau (p \to \pi^0 e^+)$ [years], the green band corresponds to $\Omega_\chi h^2 = 0.1164 - 0.1236$. The grey and orange shaded regions are those favored by the Planck and P-ACT-LB datasets, respectively. 
} 
  \label{fig:pgm}
\end{figure}

\begin{figure}
  \centering
  \subcaptionbox{\label{fig:pgm50_spt} $N_* = 50$ }{
  \includegraphics[width=0.48\columnwidth]{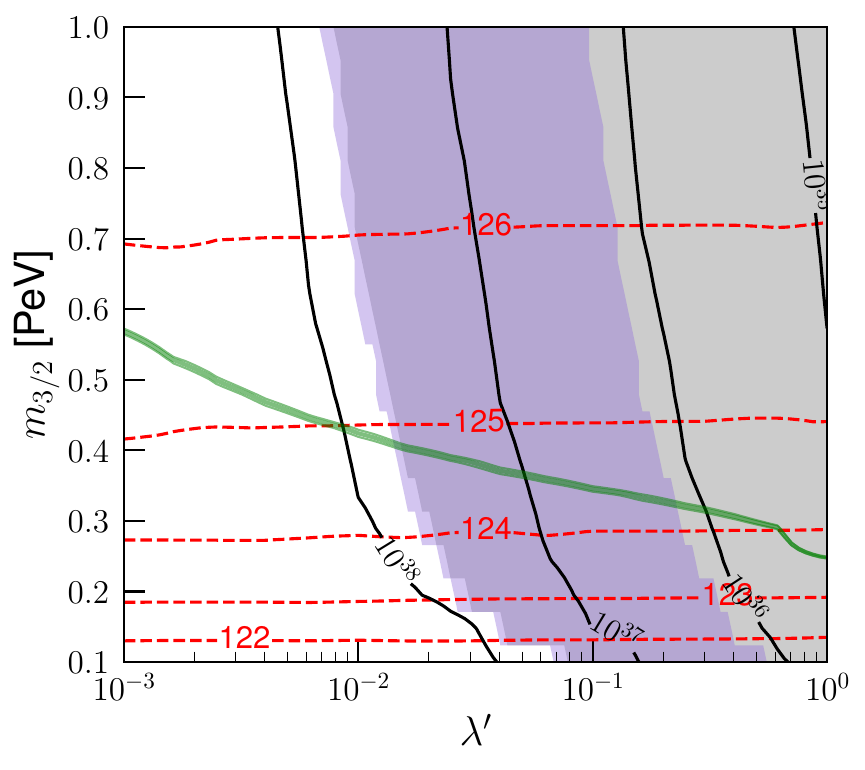}}
  \subcaptionbox{\label{fig:pgm60_spt} $N_* = 60$}{
  \includegraphics[width=0.48\columnwidth]{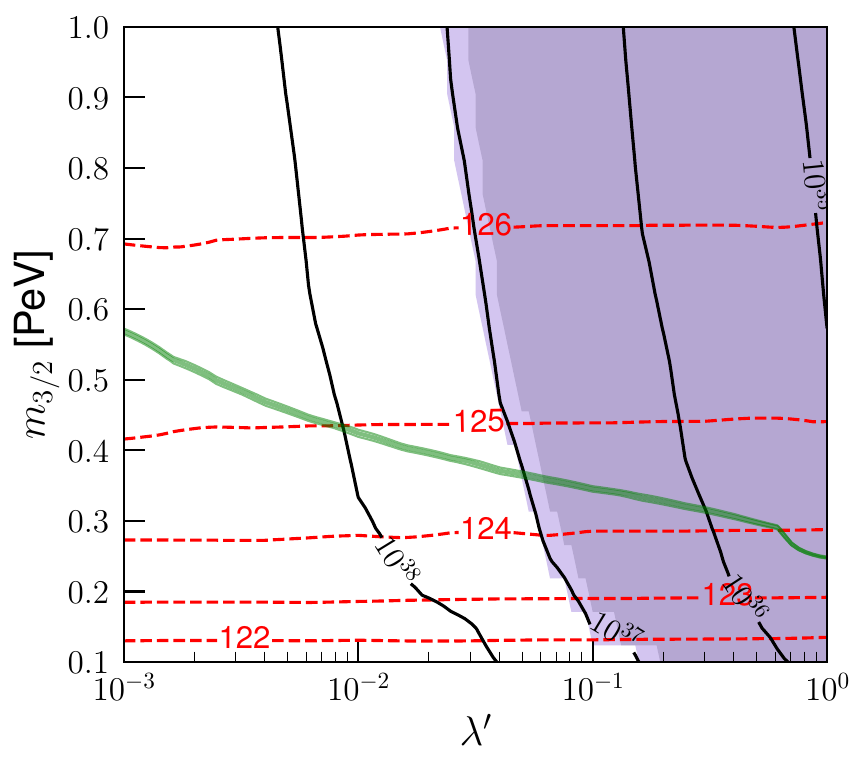}}
\caption{ As in Fig.~\ref{fig:pgm}, but using the P-ACT-SPT combination of data. 
} 
  \label{fig:pgm_spt}
\end{figure}

A broader view of the parameter space of minimal supersymmetric SU(5) with PGM can be seen in Figs.~\ref{fig:pgm} and \ref{fig:pgm_spt}, which show the $(\lambda',m_{3/2})$ planes for fixed $M_{\rm in} = 10^{18}$~GeV, $\lambda = 1$, $\tan \beta = 3$ and $\mu < 0$ for the choices $N_* = 50$ (left panel) and $N_* = 60$ (right panel). The green shaded strip shows the area where the relic density is within $3\sigma$ of the Planck determined cold dark matter density ($\Omega_\chi h^2 = 0.1164 - 0.1236$). The nearly horizontal red dot-dashed lines correspond to contours of constant Higgs mass labeled in GeV. The 
black solid curves are contours of the lifetime for $p \to \pi^0 e^+$ decay. The grey and orange shaded regions where the predictions for $n_s$ lie within the ranges favored by the Planck and P-ACT-LB datasets, in Fig.~\ref{fig:pgm}, respectively. Similarly, the grey and purple shaded regions show the favored ranges  for the Planck data and for the P-ACT-SPT combination in Fig~\ref{fig:pgm_spt}, respectively. We see that the bands in these regions that are compatible with the cosmological CDM density also predict values of $m_h$ that are compatible with the experimental measurement of the Higgs mass, within theoretical uncertainties, as well as ranges of the lifetime for $p \to \pi^0 e^+$ decay that are well beyond the current experimental limit.

\section{Update on SO(10)}
\label{sec:SO10}

In this Section, we consider an update of a no-scale GUT scenario based on the SO(10) gauge group \cite{egnno1}, whose symmetry is broken according to the following chain:
\begin{equation}
 {\rm SO}(10) \xrightarrow[{\bf 210}]{} 
 G_{\rm int} \xrightarrow[{\bf 16}, \overline{\bf 16}]{}
 G_{\rm SM} \xrightarrow[H_u, H_d]{}
 {\rm SU}(3)_C \otimes {\rm U}(1)_{\rm EM} ~.
\label{eq:symbrchain}
\end{equation}
The {\bf 210} field that breaks SO(10) at the GUT scale is denoted by $\Sigma$. The {\bf 16} and $\overline{\bf 16}$ representations that break the theory to the MSSM are denoted by $\Phi$ and $\bar{\Phi}$, respectively, and $H$ is the {\bf 10} representation of SO(10) whose SU(2)$_L$ doublet components mix with $\Phi$ and $\bar{\Phi}$ to yield the MSSM Higgs fields $H_u$ and $H_d$. The MSSM matter multiplets are in the {\bf 16} representation and identified as $\psi_i$ (where $i=1,2,3$ is the generation index), and $\phi_i$ $(i=1,2,3)$ denote SO(10) singlet ({\bf 1}) superfields, one of which is to be
identified as the inflaton,  $\phi$. Omitting the tensor structure and the generation indices, the superpotential is taken to be
\begin{align} \notag
W &= \frac{M}{2}\phi^2 - \frac{\lambda_\phi}{3\sqrt{3}} \phi^3 + y H\psi\psi + (M+b\phi)\bar{\Phi}\psi + m_{\Phi}\bar{\Phi}\Phi + \frac{\eta}{4!}\bar{\Phi}\Phi\Sigma + \frac{m_{\Sigma}}{4!}\Sigma^2 + \frac{\Lambda}{4!}\Sigma^3 \\[5pt] 
&+ m_{H}H^2 + \lambda_{\phi H} \phi H^2
+H(\alpha\Phi\Phi+\bar{\alpha}\bar{\Phi}\bar{\Phi}
+ \alpha^\prime \Phi \psi
) + c\phi \bar{\Phi}\Phi + \frac{b^\prime}{4!}\bar{\Phi}\psi\Sigma + \frac{\gamma}{4!}\phi \Sigma^2 + \kappa \,  \label{Wgen}.
\end{align}
Each of these superpotential terms plays a specific phenomenological role. 
The first two terms are responsible for producing the Starobinsky potential 
as in Eqs.~(\ref{WZW}) and (\ref{W5}). The third term determines the Standard Model Yukawa couplings.
 The fourth, tenth, and 12th terms include couplings between the
inflaton $\phi$ and Standard Model fields. Note that the magnitudes 
of these couplings determine
the neutrino masses and the decay rate of the inflaton. The Standard Model singlet
components of $\Phi$, $\bar{\Phi}$, and $\Sigma$ can acquire
non-vanishing vevs through the couplings included in the fifth through
eighth terms. After these fields develop vevs, the $\alpha H\Phi \Phi$
and $\bar{\alpha} \bar{H} \bar{\Phi} \bar{\Phi}$ terms induce mixing among the
SU(2)$_L$ doublet components inside $H$, $\Phi$, and $\bar{\Phi}$, and
by appropriately choosing these couplings we can make two linear
combinations of these fields, denoted by $H_u$ and $H_d$, much lighter
than the GUT and intermediate scales \cite{Sarkar:2004ww}, thereby realizing the
necessary doublet-triplet splitting.  
In addition, after $\Phi$ acquires a vev, the $\alpha^\prime H\Phi \psi$
term induces an $R$-parity-violating term $H_u L$, where $L$ is the
SU(2)$_L$ doublet lepton field. The remaining terms are all allowed by gauge invariance.

The SO(10) no-scale K\"ahler potential is taken to be 
\beq\label{Kgen}
K=-3\ln\left[T+T^*-\frac{1}{3}\left(\phi^*\phi + H^{\dagger}H +
\psi^{\dagger}\psi + \Phi^{\dagger}\Phi + \bar{\Phi}^{\dagger}\bar{\Phi}
+ \frac{1}{4!}\Sigma^{\dagger}\Sigma \right)\right]\,,
\eeq
and the constant $\kappa$ in (\ref{Wgen}) is introduced to yield a weak-scale gravitino mass through the relation $m_{3/2}=\langle e^{K/2}W\rangle$. All fields are taken to be untwisted. 

In order to parametrize the vevs responsible for the symmetry breaking, it is convenient to express the component fields in terms of the ${\rm SU}(4)_C\otimes {\rm SU}(2)_L \otimes {\rm SU}(2)_R$ quantum numbers:
\begin{align}
 p &= \langle \Sigma ({\bf 1}, {\bf 1}, {\bf 1}) \rangle ~, ~~~~~~
 a = \langle \Sigma ({\bf 15}, {\bf 1}, {\bf 1}) \rangle ~, ~~~~~~
 \omega= \langle \Sigma ({\bf 15}, {\bf 1}, {\bf 3}) \rangle ~,
 \nonumber \\
 \phi_R &= \langle \Phi (\overline{\bf 4}, {\bf 1}, {\bf 2}) \rangle ~,
~~~~~~
 \bar{\phi}_R = \langle \bar{\Phi} ({\bf 4}, {\bf 1}, {\bf 2}) \rangle ~,
~~~~~~
 \widetilde{\nu}_R = \langle \psi (\overline{\bf 4}, {\bf 1}, {\bf 2})
 \rangle ~.
\label{eq:vevs}
\end{align}
In terms of these singlet fields and the inflaton $\phi$, the superpotential can be simplified to 
\begin{align}\notag
W &=  \frac{M}{2}\phi^2 
- \frac{\lambda_\phi}{3\sqrt{3}} \phi^3 
 - (M+ b \phi)\bar{\phi}_R\nu_R 
+ (\eta\phi_R+ b^\prime\nu_R) \bar{\phi}_R(p+3a+6\omega) \\
&-(m_{\Phi} + c\phi) \bar{\phi}_R\phi_R
+ (m_{\Sigma}+\gamma \phi)(p^2+3a^2+6\omega^2) +2\Lambda(a^3+3p\omega^2+6a\omega^2) + \kappa ~.
\label{superpot}
\end{align}
assuming that the rest of the fields have vanishing vevs. Substitution into the scalar potential 
\beq
V=e^{2K/3}|W^i|^2\,,
\eeq
and requiring that the vevs of $p$, $a$, $\omega$,
$\phi_R$, and $\bar{\phi}_R$ do not break supersymmetry, we obtain the following set of equations that must be satisfied in a symmetry-breaking minimum:
\begin{align}
 2m_{\Sigma}p + 6\Lambda \omega^2 + \eta \phi_R\bar{\phi}_R &=0\,,
 \label{eq:vevp}\\ 
2m_{\Sigma}a + 2\Lambda (a^2+2\omega^2) + \eta \phi_R\bar{\phi}_R &=0\,,
\label{eq:veva}\\
2m_{\Sigma}\omega + 2\Lambda(p+2a) \omega + \eta \phi_R\bar{\phi}_R
 &=0 \label{eq:vevw}\,, \\
\bar{\phi}_R \left[-m_{\Phi}+\eta(p+3a+6\omega)\right] &= 0\,, \label{eq:vevphi}\\
\phi_R \left[-m_{\Phi}+\eta(p+3a+6\omega)\right] &= 0\,, \label{eq:vevphibar}\\
-c\phi_R\bar{\phi}_R + \gamma(p^2+3a^2+6\omega^2) &=0 \label{eq:vevsing}\,,\\ 
\bar{\phi}_R\left[-M+b^\prime(p+3a+6\omega)\right] &= 0\label{eq:vevnur}\,.
\end{align}
We require that the vevs of $\phi$ and $\widetilde{\nu}_R$ vanish. The solution $p=a=\omega=\phi_R=\bar{\phi}_R=0$ corresponds to the SO(10)-preserving vacuum. Non-trivial solutions are conventionally parametrized as follows~\cite{Bajc:2004xe,Aulakh:2003kg}:~\footnote{Note that a typo in the expression for the vev of $\phi_R$ given in~\cite{egnno1} has been corrected. This does not affect any of the results therein.}
\begin{alignat}{2} 
& p= -\frac{m_{\Sigma}}{\Lambda}\,\frac{y(1-5y^2)}{(1-y)^2}\,,
 \qquad\quad &&
 a=-\frac{m_{\Sigma}}{\Lambda}\,\frac{1-2y-y^2}{1-y}\,, \nonumber\\
& \omega= \frac{m_{\Sigma}}{\Lambda}y\,,   && \phi_R^2 =
 \frac{2m_{\Sigma}^2}{\eta\Lambda}\,\frac{y(1-3y)(1+y^2)}{(1-y)^2} 
\label{omphi} \,,
\end{alignat}
where $|\phi_R|=|\bar{\phi}_R|$ to ensure the vanishing of the $D$-terms, and where $y$ is a solution of the cubic equation
\beq\label{xcubic}
8y^3-15y^2+14y-3 = (y-1)^2\frac{\Lambda m_{\Phi}}{\eta m_{\Sigma}}\,.
\eeq
In order to realize Starobinsky-like inflation, we take $c=\gamma=0$ and $\lambda_\phi=M/M_P$, and assume that the symmetry-breaking fields are displaced from their vacuum values during inflation by a small amount, as justified in~\cite{egnno1}. When this is the case, the scalar potential during inflation can be simplified to
\beq \label{Vsimple}
V\simeq \frac{|M \phi - \frac{1}{\sqrt{3}} \lambda_\phi \phi^2|^2+|\phi|^2|b \phi_R |^2}{\left[1-\frac{1}{3}(|\phi|^2 + |p|^2 + 3|a|^2 + 6|\omega|^2 +2|\phi_R|^2  )\right]^2} \, .
\eeq
Equivalently, introducing the canonically-normalized inflaton field $x$ as in Eq.~(\ref{canphi})
and the parameter
\beq
\Delta K \equiv |p|^2 + 3|a|^2 + 6|\omega|^2 + 2|\phi_R|^2\,,
\label{deltak}
\eeq
the potential takes the form
\begin{align}\notag
V &= \left(1-\tanh ^2(x/\sqrt{6})-\tfrac{1}{3}\Delta K \right)^{-2} \\ \label{Vfull}
&\quad\qquad \times 3 \tanh ^2 (x/\sqrt{6} ) \left[ M^2\left(\tanh (x/\sqrt{6} ) -1\right)^2 + |b\phi_R|^2  \right]\\ \label{Vnotfull}
&\simeq \frac{3}{4}M^2\left(1-e^{-\sqrt{2/3}\,x}\right)^2 +\Delta V\,,
\end{align}
where
\beq\label{deltaV}
\Delta V = \left[\frac{3}{4} |b\phi_R|^2 + \frac{1}{2}M^2 e^{-\sqrt{2/3}\,x}\Delta K \right]  \sinh ^2 (\sqrt{2/3}\, x )\,.
\eeq\par
For any given GUT-breaking field configuration, the shape of the potential, and thus the CMB observables, will be determined by the values of the inflaton mass $M$ and the coupling $b$. We limit our numerical studies here to two realistic cases~\cite{egnno1} in which we determine these dependencies in a self-consistent manner.\\

\begin{figure}[ht!]
\centering
\includegraphics[width=\textwidth]{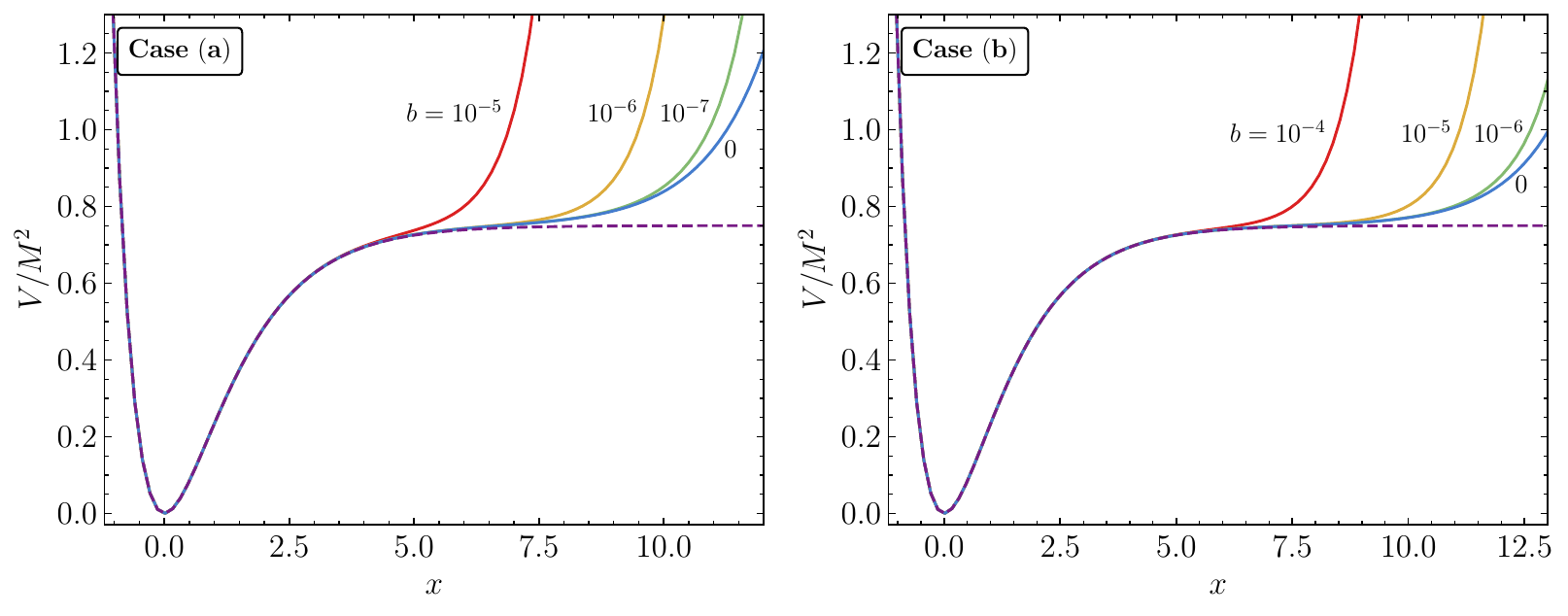}
\caption{ The inflationary potential (\ref{Vfull}) for the SO(10) GUT for two realizations of the symmetry-breaking pattern, and selected values of the coupling $b$. Left: Case (a), with $M_{\rm int}=M_{\rm GUT}$ and $\Delta K\simeq 2\times 10^{-4}$. Right: Case (b), with $M_{\rm int}<M_{\rm GUT}$ and $\Delta K\simeq 4.8\times 10^{-5}$. In both panels the dashed line corresponds to the Starobinsky potential.}
 
  \label{fig:pot10}
\end{figure}

\begin{figure}[ht!]
\centering
\includegraphics[width=\textwidth]{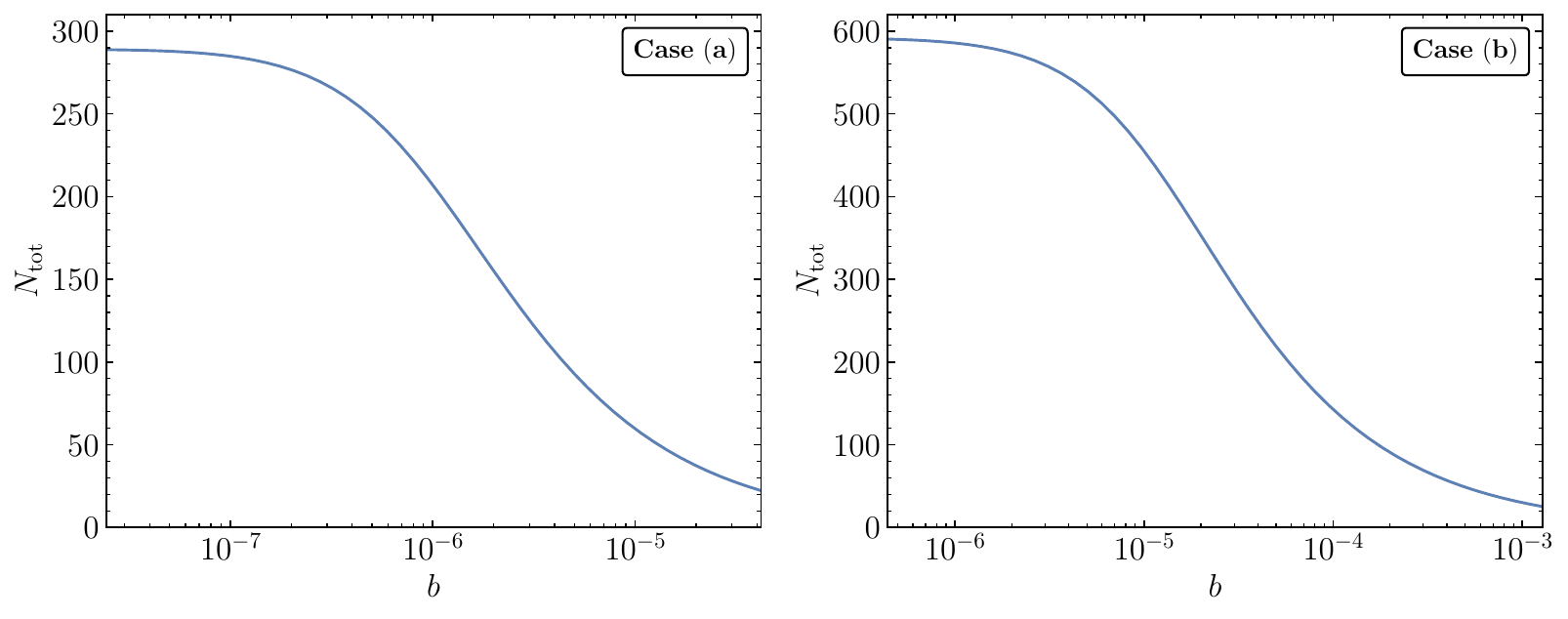}
\caption{ Total duration of inflation, $N_{\rm tot}$, as a function of the coupling $b$ for the two realizations of the SO(10) GUT shown in Fig.~\ref{fig:pot10}. The initial condition is taken as the value of $x$ such that $V/M^2=2$. }
  \label{fig:Ntot10}
\end{figure}

\begin{figure}[ht!]
    \centering
    \includegraphics[width=\textwidth]{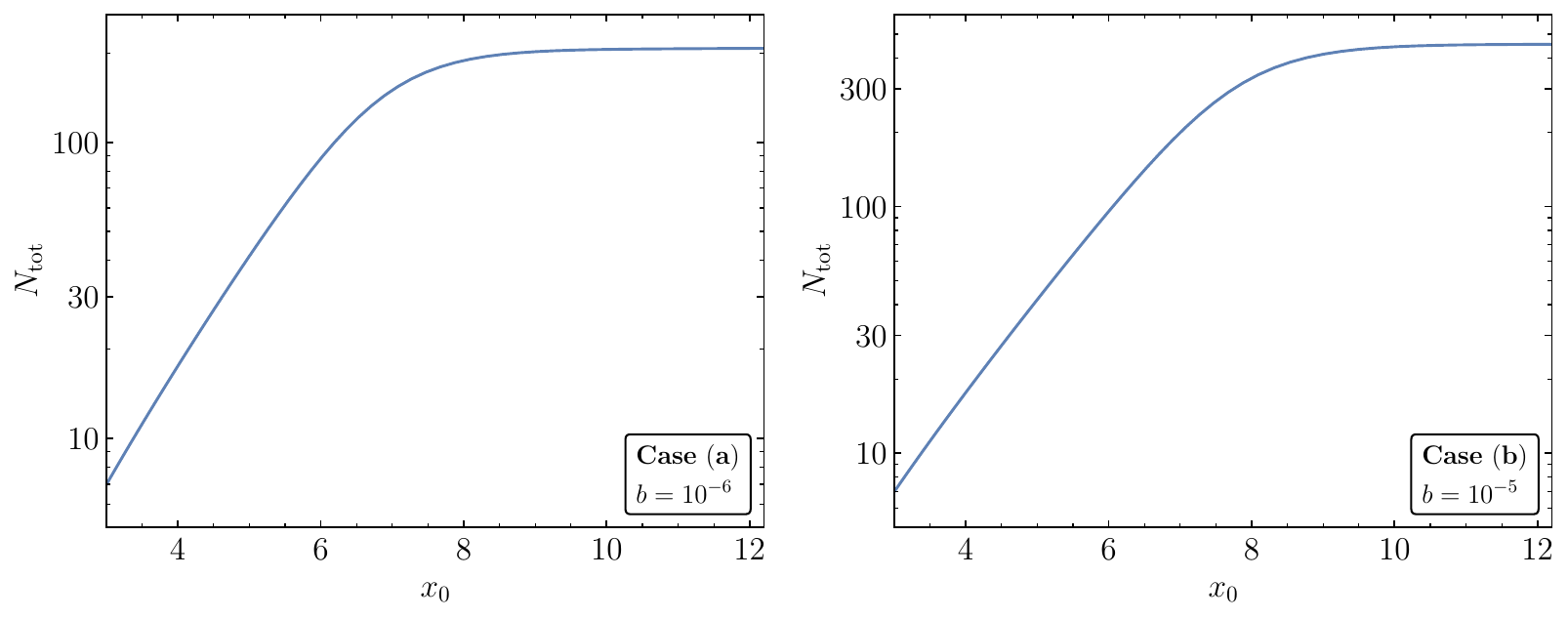}
    \caption{ Total duration of inflation, $N_{\rm tot}$, as a function of the inflaton initial condition $x_0$, for the two realizations of the SO(10) model, and fixed $b$.}
    \label{fig:Ntotx}
\end{figure}

{\bf Case (a)}: The SO(10) gauge group is broken directly into the SM gauge group at the GUT scale, $M_{\rm int}=M_{\rm GUT}$. This scenario can be realized with $y= -1$ and $\phi_R = p =a = \omega = M_{\rm GUT}=10^{16}\ {\rm GeV}$, corresponding to the parameters
\beq\label{partpar}
m_{\Phi}= 3.3\times 10^{-2}\,, \quad m_{\Sigma}= 8.2\times 10^{-4}\,,\quad  \Lambda=-0.2\,,\quad  \eta=0.8\,.
\eeq
This choice fixes $\Delta K\simeq 2\times 10^{-4}$, leaving only $b$ and $M$ as free parameters in Eq.~(\ref{deltaV}). The corresponding form of the inflaton potential is displayed in the left panel of Fig.~\ref{fig:pot10}. As is shown there, the flatness of the potential is lost at large field values, with larger values of $b$ restricting the duration of inflation to shorter ranges. The total duration of inflation, $N_{\rm tot}$ is shown as a function of $b$ in the left panel of Fig.~\ref{fig:Ntot10} taking an initial field value corresponding to $V/M^2 = 2$. In order to have $N_{\rm tot}\geq50$ (60), one must require $b\lesssim 1.3\times 10^{-5}$ ($10^{-6}$). We note that $N_{\rm tot}\lesssim 290$, as even $b=0$ leads to an exponentially steep slope at large inflaton values, since $\Delta K\neq 0$. This conclusion is not affected by the choice of the initial field value, as larger initial values of $x$ would lead initially to a fast evolution of the inflaton slowed by Hubble friction leaving the total number of e-folds largely independent of the initial field value. This is shown explicitly in the left panel of Fig.~\ref{fig:Ntotx}, where we see that even for very large initial field values, $N_{\rm tot} \lesssim 200$ for $b=10^{-6}$.  This is similar to the deformation seen earlier for SU(5) for non-zero values of $V_\Sigma$ (in Fig.~\ref{staroN}) and similar to the no-scale deformation of the Starobinsky model discussed in \cite{Antoniadis:2025pfa}.   \\

{\bf Case (b)}: The SO(10) gauge group is broken first into an intermediate group, $M_{\rm int} < M_{\rm GUT}$. This scenario can be realized with $\phi_R < p, a, \omega$, corresponding to $y\simeq 0,\, 1/3$ or $\pm i$~\cite{Bajc:2004xe}. We consider $y\ll 1$, for which $G_{\rm int}={\rm SU(3)}_C\otimes {\rm SU(2)}_L\otimes {\rm SU(2)}_R\otimes {\rm U(1)}_{B-L}\otimes D$, where $D$ denotes $D$-parity~\cite{Kuzmin:1980yp,Kibble:1982dd,Chang:1983fu,Chang:1984uy,Chang:1984qr}. Specifically, we  realize this scenario numerically by taking $y=0.0004$ and 
\beq\label{partparb}
m_{\Phi}= 1.2\times 10^{-3}\,, \quad m_{\Sigma}= 4\times 10^{-4}\,,\quad  \Lambda=-0.1\,,\quad  \eta=0.1\,.
\eeq
This corresponds to the vevs $a\simeq 9.7\times 10^{15}\,{\rm GeV}$, $\phi_R\simeq 2.8\times 10^{14}\,{\rm GeV}$ and $p\simeq -\omega\simeq 4\times 10^{12}\,{\rm GeV}$, and $\Delta K\simeq 4.8\times 10^{-5}$. The form of the inflaton potential (\ref{deltaV}) as $b$ is varied is shown in the right panel of Fig.~\ref{fig:pot10}. The corresponding total number of e-folds as a function of $b$ is shown in the right panel of Fig.~\ref{fig:Ntot10}. In this case $N_{\rm tot}\geq50$ (60) requires $b\lesssim 4.8\times 10^{-4}$ ($3.7\times 10^{-4}$). The maximum duration of inflation, $N_{\rm tot}\simeq 590$, is larger than in case (a) due to the smaller value of $\Delta K$.  Nevertheless, the total number of e-folds remains independent of the initial conditions as seen in the right panel of Fig.~\ref{fig:Ntotx} for $b= 10^{-5}$. \\

Alternative realizations of the intermediate gauge group include $G_{\rm int}={\rm SU(5)}\otimes {\rm U(1)}$ for $y\simeq 1/3$. This choice leads to rapid proton decay, because the masses of the SU(5) gauge bosons are $\mathcal{O}(M_{\rm int})$. For $y\simeq \pm i$, one can show that $G_{\rm int}={\rm SU(3)}_C\otimes {\rm SU(2)}_L\otimes {\rm U(1)}_R\otimes {\rm U(1)}_{B-L}$. The realization of Starobinsky-like inflation in this case leads to similar phenomenology as case (b) above.\\

In the slow-roll approximation, the CMB observables can be evaluated by means of Eqs.~(\ref{epsilon})-(\ref{r}), with the amplitude of the scalar spectrum given by Eq.~(\ref{eq:As}). For $b\phi_R,\Delta K \ll 1$ an analytic approximation is available:
\begin{align}
n_s \;&=\; -\frac{2}{N_*} + \frac{8}{3}\left(\frac{b\phi_R}{M}\right)^2N_*^2 + \frac{32}{81}\Delta K\ N_*\,,\\
r\;&=\; \frac{12}{N_*^2} + \frac{32}{3} \left(\frac{b\phi_R}{M}\right)^2 N_* + \frac{64}{27}\Delta K\,.
\end{align}
However, in order to evaluate the CMB observables exactly we compute numerically the primordial scalar and tensor power spectra by solving the equations of motion for the corresponding gauge-invariant fluctuations.~\footnote{The full details of this computation can be found in Appendix B of~\cite{egnov} and Section 2.2 of~\cite{Garcia:2024rwg}.}   Analytical expressions for the power spectra parameters become cumbersome for $b,\Delta K\neq0$, while a numerical evaluation is straightforward. In particular, the number of e-folds $N_*$ depends on the duration of the reheating epoch, which is determined by $b$. 

As discussed in detail in~\cite{egnno1}, the dominant inflaton decay channels correspond to Higgs-slepton and higgsino-lepton final states, induced by $\bar{\Phi}$-$H$ and neutrino-singlino mixings, with a combined rate
\beq
\Gamma(S\rightarrow H_u\tilde{L}) + \Gamma(S\rightarrow \tilde{H}_u L) =
\frac{M}{4\pi}\left|C_{SHL}\right|^2\,,
\eeq
where
\begin{equation}
C_{SHL} =  b\left({\cal U}_{21} -\frac{f_\nu\phi}{M}\right) ~.
\end{equation}
With ${\cal U}_{21} ={\cal O}(1)$, $f_\nu ={\cal O}(10^{-5})$ and $\phi/M \lesssim{\cal O}(10^3)$ in our setup, the decay of the inflaton is determined by the magnitude of the coupling $b$. Therefore, as mentioned above, the number of e-folds depends on $b$ not only through $V_*$, but also through the reheating dependent term (\ref{eq:nstargamma}). 

While we have not performed a detailed study of proton decay in this model we do note the following. The dominant decay mode is expected to be $p \to K^+ {\bar \nu}$. The decay amplitude is enhanced by wino and higgsino exchange proportional to $1/\sin 2\beta \simeq (\tan \beta)/2$ which is favoured to be large in SO(10) models. In addition, if the intermediate scale $M_{\rm int} \simeq \phi_R \ll M_{\rm GUT}$,
the colour-triplet Higgs masses are of order the intermediate scale further enhancing the proton decay rate. To suppress the rate, it is necessary to consider a relatively high supersymmetry breaking scale and/or some mechanism to suppress the colour-triplet contribution. Thus models where the intermediate scale is close to the GUT scale are preferable. 
Case a) with $M_{\rm int} = M_{\rm GUT}$ is one such example.

\begin{figure}[t!]
\centering
\includegraphics[width=0.60\columnwidth]{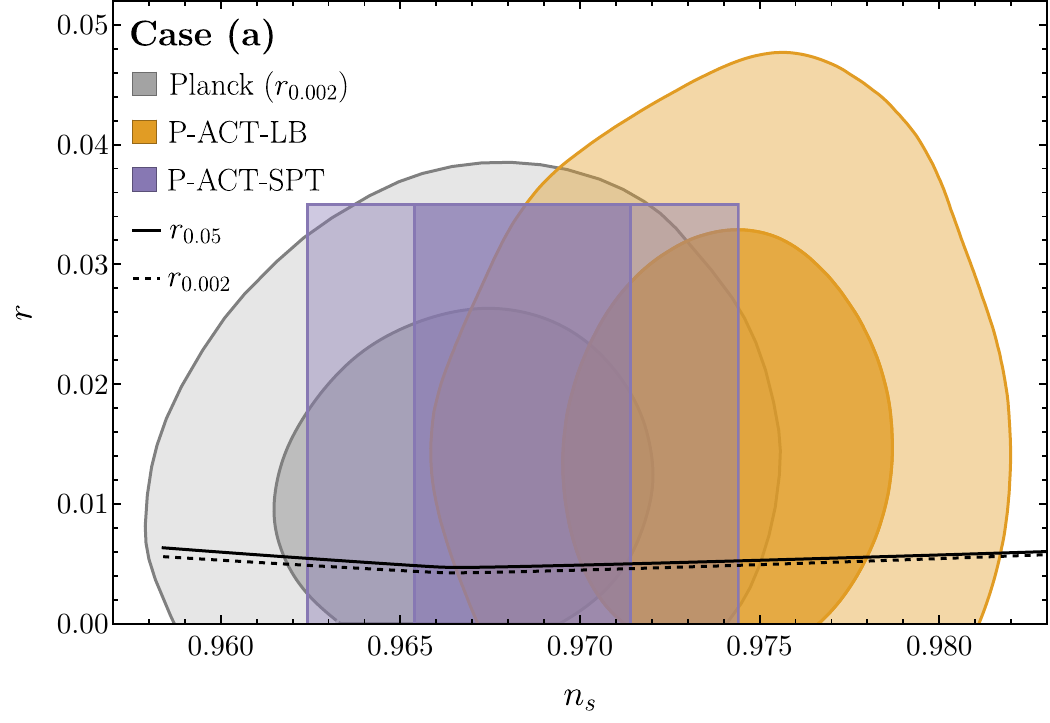}\\[5pt]
\includegraphics[width=0.49\columnwidth]{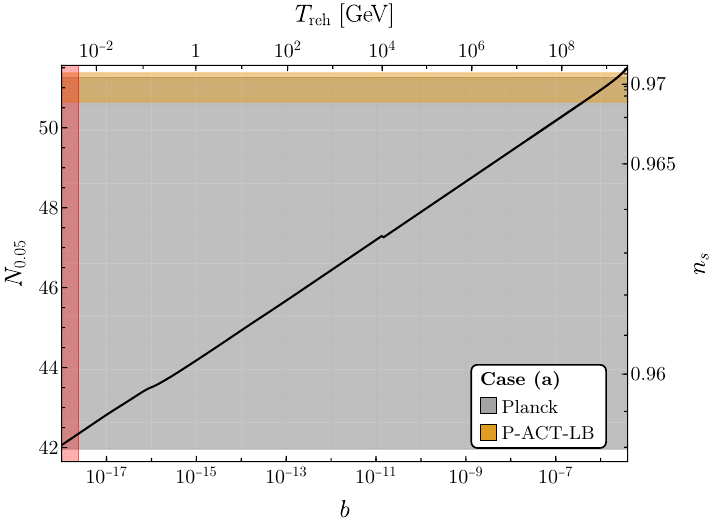} \includegraphics[width=0.49\columnwidth]{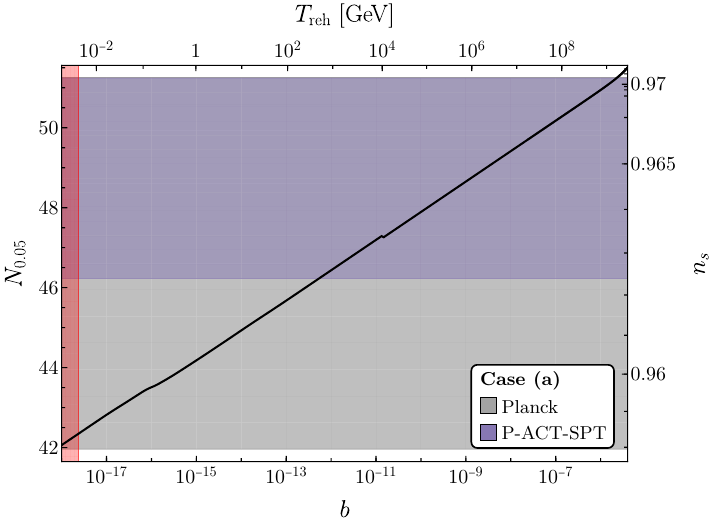}
\caption{ {\it Upper panel}: CMB observables for case (a) of the SO(10) model, compared to the Planck $1\sigma$ and $2\sigma$ contours (gray), the P-ACT-LB results (orange), and the P-ACT-SPT results for $n_s$ only, for $2.5\times 10^{-18}\lesssim b\lesssim 4.4\times 10^{-6}$. The continuous black curve corresponds to the evaluation of the tensor-to-scalar ratio at the pivot scale $k_*=0.05\,{\rm Mpc}^{-1}$, as used by the ACT analysis, while the dashed line corresponds to the evaluation of $r$ at the WMAP pivot scale $k_*=0.002\,{\rm Mpc}^{-1}$, for proper comparison with the Planck results. {\it Lower panel}: Values of $N_*$ and $n_s$ as functions of $b$, or equivalently the reheating temperature. The shaded regions correspond to those in the upper panel. }
 
  \label{fig:ca10}
\end{figure}

The upper panel of Fig.~\ref{fig:ca10} shows the CMB observables in the $(n_s,r)$ plane for case (a) of the SO(10) model. These are compared to the regions allowed by the Planck data at the
$1$ and $2\sigma$ levels (grey), the P-ACT-LB results (orange) and the P-ACT-SPT combination for $n_s$ only (purple), for $2.5\times 10^{-18}\lesssim b\lesssim 4.4\times 10^{-6}$. The lower bound on $b$ corresponds to a lower bound on the reheating temperature of 4 MeV, while the upper bound corresponds to a scale-invariant scalar spectrum, $n_s=1$, with $N_*\simeq51.6$.  Similarly to Fig.~\ref{fig:cmbsu5}, the continuous black curve corresponds to the evaluation of the tensor-to-scalar ratio at the pivot scale $k_*=0.05\,{\rm Mpc}^{-1}$ as assumed in the Planck/ACT analysis, while the dashed line corresponds to the evaluation of $r$ at the WMAP pivot scale $k_*=0.002\,{\rm Mpc}^{-1}$ as assumed in the Planck analysis. We see that the results for the CMB observables are very similar. The lower panels of Fig.~\ref{fig:ca10} show the values of $N_*$ and $n_s$ as functions of $b$, or equivalently the reheating temperature. The shaded bands correspond to the Planck and P-ACT-LB (left) or P-ACT-SPT (right) regions in the upper panel. We see that the range of $n_s$ favored by the Planck analysis corresponds to lower values of $b$ and the reheating temperature than those favored by the P-ACT-LB analysis.

\begin{figure}[t!]
\centering
\includegraphics[width=0.60\columnwidth]{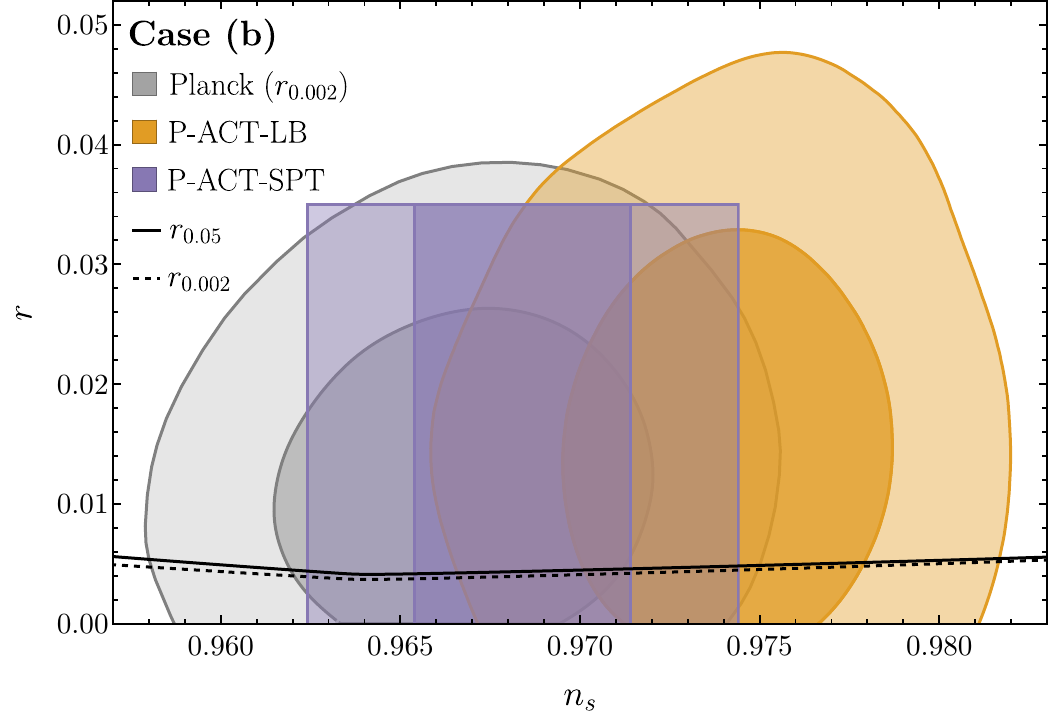}\\[5pt]
\includegraphics[width=0.49\columnwidth]{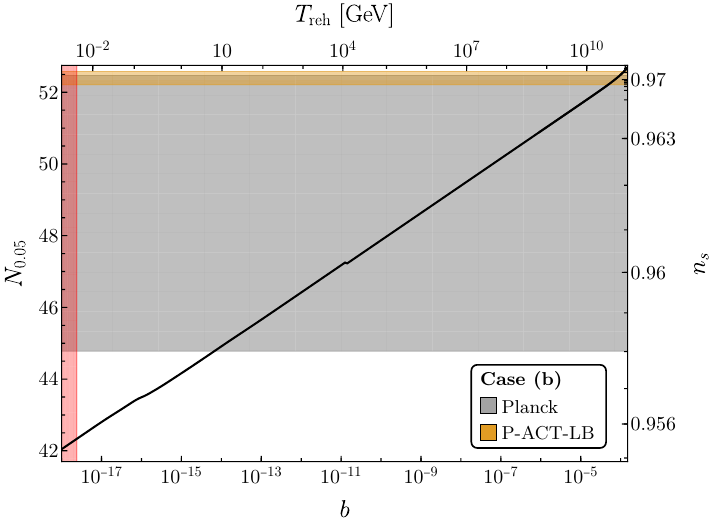} \includegraphics[width=0.49\columnwidth]{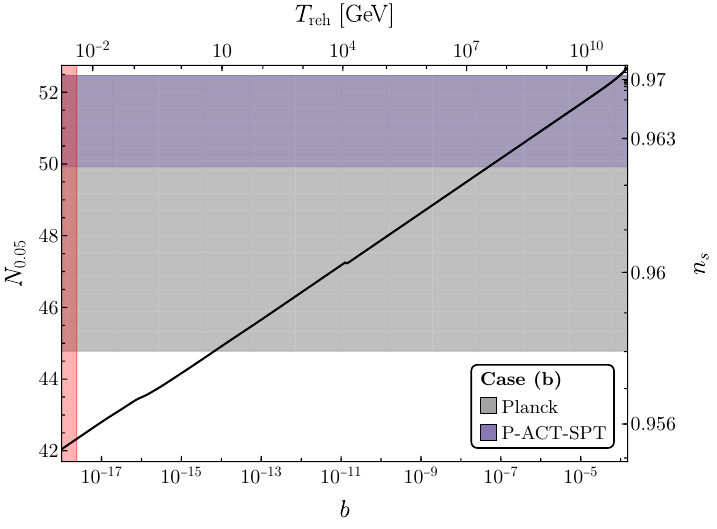}
\caption{ {\it Upper panel}: CMB observables for case (b) of the SO(10) model, compared to the combined Planck $1\sigma$ and $2\sigma$ contours (gray), the P-ACT-LB results (orange), and the P-ACT-SPT data for $n_s$ only (purple), for $2.5\times 10^{-18}\lesssim b\lesssim 1.6\times 10^{-4}$. The lines and shadings correspond to those in Fig.~\ref{fig:ca10}. 
} 
  \label{fig:cb10}
\end{figure}

Fig.~\ref{fig:cb10} shows the corresponding results for case (b) of the SO(10) model, and we see that the results are quite similar. We conclude that both the SO(10) cases studied can comfortably accommodate any of the Planck, P-ACT-LB and P-ACT-SPT values of the CMB observables.

\section{Summary and Conclusions}
\label{sec:conx}

Embedding an inflation model in a more complete UV theory is always a challenge. 
As is well known, minimal supergravity formulations of inflation typically lead to the eta-problem \cite{eta}. This problem can be alleviated in no-scale supergravity \cite{nsi}. Moreover, no-scale supergravity accommodates relatively simple derivations \cite{eno6,eno7,Ellis:2020lnc} of the Starobinsky model \cite{Staro}, as well as its attractor variants \cite{Kallosh:2013yoa,enov3,enov4,Garcia:2020eof}.
Nevertheless, in theories with additional fields, even no-scale supergravity presents challenges, since the K\"ahler field space manifold is complex and leads to mixing between the fields in a canonical basis. This is readily apparent in both the SU(5) and SO(10) grand unified theories discussed here.

The mixing in the scalar potential invariably leads to deformations away from the flatness of the Starobinsky potential at large field values. These generally have a strong impact on the tilt of the scalar fluctuation spectrum, but less of an effect on the tensor-to-scalar ratio.

In this work we considered two embeddings of GUTs in the no-scale formulation of the Starobinsky model.  One is based on SU(5), the other on SO(10). In SU(5), the deformation of the potential, $\Delta V$ is given by Eq.~(\ref{eq:delvsu5}) and is proportional to the square of the vev of the SU(5) adjoint Higgs field, $V_\Sigma^2$. This deformation increases systematically the value of $n_s$, not only improving the compatibility with the Planck determination of $n_s$, but also allowing concordance with the determination made by ACT using Planck and BAO data. The experimental constraints on $n_s$ may be used to bound the Higgs adjoint vev as derived in Eqs.~(\ref{eq:limonvsigma}--\ref{eq:limonvsigma_spt}). We find that not all supersymmetric SU(5) models are equally viable. In the minimal model and in the absence of fine-tuning the $\lambda_{\phi\Sigma}$ coupling, we find that limits on the proton lifetime  and limits to $n_s$ lead to incompatible values of the Higgs adjoint trilinear superpotential coupling, $\lambda'$. In contrast, a similar model with PGM boundary conditions predicts much longer proton lifetimes, allowing larger values of $\lambda'$ and compatibility with $n_s$.

We have also made similar analyses of two SO(10) GUT models distinguished by their symmetry breaking patterns. These were previously considered in \cite{egnno1}
and have been updated here in connection with recent CMB data. Both of these models exhibit deformations similar to that observed for SU(5). The magnitudes of these deformations are related to a parameter, $b$, characterizing the superpotential mixing of the inflaton with the Higgs and matter {\bf 16}-plets that play roles in the neutrino mass matrix \cite{egnno1} and determine the reheating temperature after inflation. However, in these models, even in the absence of the coupling $b$, K\"ahler mixing (given by Eq.~(\ref{deltak})) still leads to deformations of the inflationary potential. 

Furthermore, we have shown that the deformations of the type arising in these GUT embeddings shield the potential from difficulties associated with large initial field values \cite{ano,Antoniadis:2025pfa}. In particular, the total number of e-folds is largely independent of the initial conditions. This is in sharp contrast to the Starobinsky model, where
the number of e-folds increases exponentially with the initial field value. The total number of e-folds can be interpreted as a measure of the probability that our patch of the Universe originated from a given initial state, with higher probabilities corresponding to larger numbers of e-folds. The deformations discussed here would indicate that very large field values are never attained, and initial field values $ x \gtrsim 6$ yield similar numbers of e-folds and hence may be interpreted as equally probable.  This alleviates one of the main complaints about Starobinsky-like models of inflation in the context of the Large Distance Conjecture.  

The Starobinsky model is in good agreement (within quoted uncertainties) with the Planck determination of $n_s$. 
On the other hand, it falls outside the 95\% CL range if the P-ACT-LB combination of data is used to determine $n_s$.  However, GUT deformations of the type discussed here affect the calculated value of $n_s$, enabling CMB measurements to be used as a probe of GUT-scale physics. Improved accuracy in the determination of CMB observables may provide one of our few experimental probes of GUT-scale physics.

\section*{Acknowledgments}

The work of J.E. was supported by the United Kingdom STFC Grant ST/T000759/1.
The work of M.A.G.G.~was supported by the DGAPA-PAPIIT grant IA100525 at UNAM, and the CONAHCYT ``Ciencia de Frontera'' grant CF-2023-I-17. N.N. acknowledge support from the Simons Foundation Targeted Grant 920184 to the Fine Theoretical Physics Institute. His work was also supported in part by the Grant-in-Aid for Scientific Research C (No.~25K07314).  
 The work of K.A.O.~was supported in part by DOE grant DE-SC0011842 at the University of Minnesota. 





\end{document}